\documentclass[notitlepage,showkeys,twocolumn,11pt,times,superscriptaddress, a4paper]{revtex4-2}
\usepackage{stix}
\usepackage[T1]{fontenc}
\usepackage[english]{babel}
\usepackage{graphicx}
\usepackage{bm} 
\usepackage[colorlinks=true,
            linkcolor=blue,
            urlcolor=blue,
            citecolor=blue]{hyperref}
\usepackage{lipsum}
\usepackage{amsmath}
\usepackage{physics}
\usepackage{booktabs}
\usepackage{geometry}
\usepackage{mathtools}
\usepackage{siunitx}

\DeclareMathOperator{\RTE}{RTE}

\begin{document}
    \title{Recurrence time entropy and weak chaos in Hamiltonian flows}

    \author{Matheus Rolim Sales}
    \email{rolim.sales.m@gmail.com}
    \affiliation{São Paulo State University (UNESP), Institute of Geosciences and Exact Sciences, 13506-900, Rio Claro, SP, Brazil}
    
    \author{Leonardo Costa de Souza}
    \affiliation{Institute of Physics, University of São Paulo, 05315-970, São Paulo, SP, Brazil}
    
    \author{Iberê Luiz Caldas}
    \affiliation{Institute of Physics, University of São Paulo, 05315-970, São Paulo, SP, Brazil}

    \author{Edson Denis Leonel}
    \affiliation{São Paulo State University (UNESP), Institute of Geosciences and Exact Sciences, 13506-900, Rio Claro, SP, Brazil}

    \author{José Danilo Szezech Jr.}
    \affiliation{Graduate Program in Science, State University of Ponta Grossa, 84030-900, Ponta Grossa, PR, Brazil}
    \affiliation{Department of Mathematics and Statistics, State University of Ponta Grossa, 84030-900, Ponta Grossa, PR, Brazil}

    \date{\today}

    \begin{abstract}
            Stickiness in mixed Hamiltonian systems causes chaotic trajectories to remain temporarily trapped near regular structures, making it difficult to distinguish regular, weakly chaotic, and strongly chaotic motion over finite times. We show that the recurrence time entropy (RTE), previously used in discrete maps, also characterizes weak chaos in Hamiltonian flows. In the Hénon–Heiles system, the RTE reproduces the structures identified by the largest Lyapunov exponent, taking intermediate values in the sticky layers, and yields a proportion of chaotic trajectories consistent with the smaller alignment index. The finite-time RTE identifies low-entropy episodes near regular islands, whose durations decay algebraically, whereas high-entropy episodes decay exponentially. The same characterization holds for a periodically driven system and for a Hamiltonian system with three degrees of freedom, whose section is four-dimensional. The RTE is thus an effective diagnostic of weak chaos and stickiness in Hamiltonian flows.
        \end{abstract}
    
    \maketitle

    \section{Introduction}

    Chaotic dynamics is investigated in a broad range of settings. In dissipative systems, the classical examples of the Lorenz~\cite{Lorenz1963} and Rössler~\cite{Rossler1976} flows and of Chua's circuit~\cite{Matsumoto1984} have been followed by a large body of work on attractors of increasing complexity, including the memristive circuits and neural networks that display multi--scroll and multi--butterfly attractors~\cite{Lin2026, Lin2026b, Chen2025, Zhang2025}. Chaos may also be transient, with trajectories behaving irregularly for a finite time before escaping the region of interest~\cite{Altmann2008, Altmann2009}. In conservative systems, which are governed by Hamilton's equations of motion and are the subject of this paper, the absence of attractors leads to a qualitatively different scenario, in which regular and chaotic motion coexist in a mixed phase space. In the integrable limit, trajectories lie on invariant tori and the motion is regular. According to the Kolmogorov--Arnold--Moser (KAM) theorem~\cite{lichtenberg2013regular}, sufficiently irrational invariant tori persist under small perturbations, while resonant tori are destroyed and give rise to elliptic and hyperbolic periodic structures, as described by the Poincaré--Birkhoff theorem~\cite{lichtenberg2013regular}. For Hamiltonian systems with two degrees of freedom, this mixed organization is conveniently visualized on a Poincaré surface of section, where regular islands coexist with chaotic components. The resulting phase space structure is typically organized hierarchically, with islands-around-islands embedded in chaotic regions and remnants of destroyed invariant tori, such as cantori, acting as partial barriers to transport~\cite{lichtenberg2013regular,Mackay1984,Mackay1984b,Meiss1986}.

    One of the most important consequences of this hierarchical organization is the phenomenon of stickiness~\cite{Contopoulos1971,Karney1983,Meiss1983,Efthymiopoulos1997,Cristadoro2008, CONTOPOULOS2008, Contopoulos2010}. Chaotic trajectories that approach the neighborhood of regular islands on the Poincaré section may remain trapped there for very long, but finite, times before returning to the chaotic component. During these trapping episodes, the orbit appears nearly regular over finite times, even though it remains chaotic asymptotically. This intermittent alternation between extended exploration of the chaotic component and temporary trapping near regular structures is one of the clearest manifestations of weak chaos in mixed Hamiltonian systems.  Experimental studies have found chaotic transport~\cite{solomon1993}, as well as sticking and unsticking of particles, in a laminar fluid flow within a rotating annulus. Furthermore, in astrophysical systems, the presence of stickiness could play a crucial role in the dynamics of asteroid or debris orbits~\cite{Contopoulos2002}.

    Then the stickiness has direct consequences for transport and for the numerical characterization of chaotic dynamics. Since sticky trajectories may spend long intervals near regular islands, finite--time quantities can be strongly affected by the local phase space structure. In particular, the convergence of Lyapunov exponents may become slow, because trapping episodes can temporarily reduce the finite--time estimate of the instability~\cite{Szezech2005}. For this reason, several complementary tools have been used to characterize weak chaos and sticky dynamics, including finite--time Lyapunov exponents~\cite{Szezech2005,Harle2007, Silva2015}, finite--time rotation numbers~\cite{Santos2019}, finite--time Hurst exponent~\cite{Borin2024, Borin2025}, recurrence--time statistics~\cite{Altmann2005, Altmann2006, Abud2013, Lozej2020}, and recurrence quantification analysis~\cite{Zou2007b,Zou2007,Palmero2022}.

    Recurrence plots provide a useful framework for this problem because they characterize the recurrence structure of trajectories directly from time series~\cite{Eckmann1987}. Given a trajectory in phase space, a recurrence plot encodes the pairs of times for which the orbit returns close to previously visited states. Different dynamical regimes produce different recurrence structures~\cite{Marwan2007, Zou2007, Zou2007b, Sales2023,Daquin2026}: regular trajectories are associated with long and coherent diagonal patterns, chaotic trajectories display shorter and more fragmented structures, and sticky trajectories typically exhibit intermediate patterns reflecting their intermittent character. These features make recurrence--based methods particularly suitable for detecting dynamical transitions and weakly chaotic behavior.

    Among recurrence--based measures, quantities derived from recurrence times are especially relevant for mixed Hamiltonian systems. In low-dimensional quasiperiodic dynamics, recurrence times are strongly constrained, whereas chaotic motion typically produces broader and more irregular recurrence--time distributions. The recurrence time entropy (RTE) was originally introduced independently of recurrence plots~\cite{Little2007} and was later shown to provide an estimate of the Kolmogorov--Sinai entropy from recurrence times~\cite{Baptista2010}. More recently, RTE-based approaches have been used to characterize weakly chaotic trajectories in two--dimensional area--preserving maps~\cite{Sales2023,Souza2024,Viana2025}, fractional maps~\cite{Borin2025}, and continuous-time systems~\cite{Gabrick2023}.

    In this work, we extend this recurrence--based characterization to Hamiltonian flows. We first use the Hénon--Heiles Hamiltonian as a paradigmatic autonomous system with two degrees of freedom and compute the RTE from successive crossings of a Poincaré surface of section. This allows us to compare the RTE directly with the largest Lyapunov exponent and to test whether recurrence--time statistics reproduce the mixed phase space structure of the system. We additionally compare the proportion of chaotic trajectories identified by the RTE with that obtained from the smaller alignment index (SALI) across different energies. We also analyze the finite--time RTE along individual chaotic trajectories, identify the phase space regions responsible for low-- and high--RTE episodes, and compute the corresponding distributions of episode durations. Finally, we consider a periodically driven one-degree-of-freedom Hamiltonian and compute the RTE from its stroboscopic map. This second example allows us to verify whether the same recurrence--based diagnostic works in a different Hamiltonian setting.

    Our results show that the RTE distinguishes regular islands from chaotic regions, detects the sticky transition layers that surround them, which we identify by the intermediate values assumed by the RTE, and captures intermittent trapping episodes along chaotic trajectories. In the Hénon--Heiles system, the RTE reproduces the main phase space structures observed through the largest Lyapunov exponent and exhibits a strong positive correlation with it. Moreover, the proportion of chaotic trajectories identified by the RTE is consistent with that obtained from the SALI across the energy range investigated. The finite--time analysis further reveals that low--RTE episodes are localized near regular islands and produce algebraic trapping--time statistics, whereas high--RTE episodes are associated with extended exploration of the chaotic component and produce exponentially decaying duration distributions. The application to the periodically driven Hamiltonian confirms that the same recurrence--time characterization is not specific to the Hénon--Heiles model.

    This paper is organized as follows. In Sec.~\ref{sec:henonheiles}, we introduce the Hénon--Heiles Hamiltonian and describe the Poincaré surface of section used in the analysis. In Sec.~\ref{sec:rte}, we recall the recurrence-plot construction and define the recurrence time entropy used throughout the paper. In Sec.~\ref{sec:hh_results}, we apply the RTE to the Hénon--Heiles system, compare it with the largest Lyapunov exponent, compare the proportion of chaotic trajectories with the SALI, and analyze finite--time RTE fluctuations and trapping--time statistics. In Sec.~\ref{sec:paradigm}, we apply the same recurrence--based analysis to a periodically driven Hamiltonian system using its stroboscopic map. In Sec.~\ref{sec:threedof}, we extend the analysis to a Hamiltonian system with three degrees of freedom, whose surface of section is four--dimensional. Finally, Sec.~\ref{sec:conclusions} summarizes our results. The dependence of the RTE on the length of the time series is examined in Appendix~\ref{sec:appendix}.

    \section{The Hénon--Heiles Hamiltonian}
    \label{sec:henonheiles}

    \begin{figure*}[t!]
        \centering
        \includegraphics[width=\linewidth]{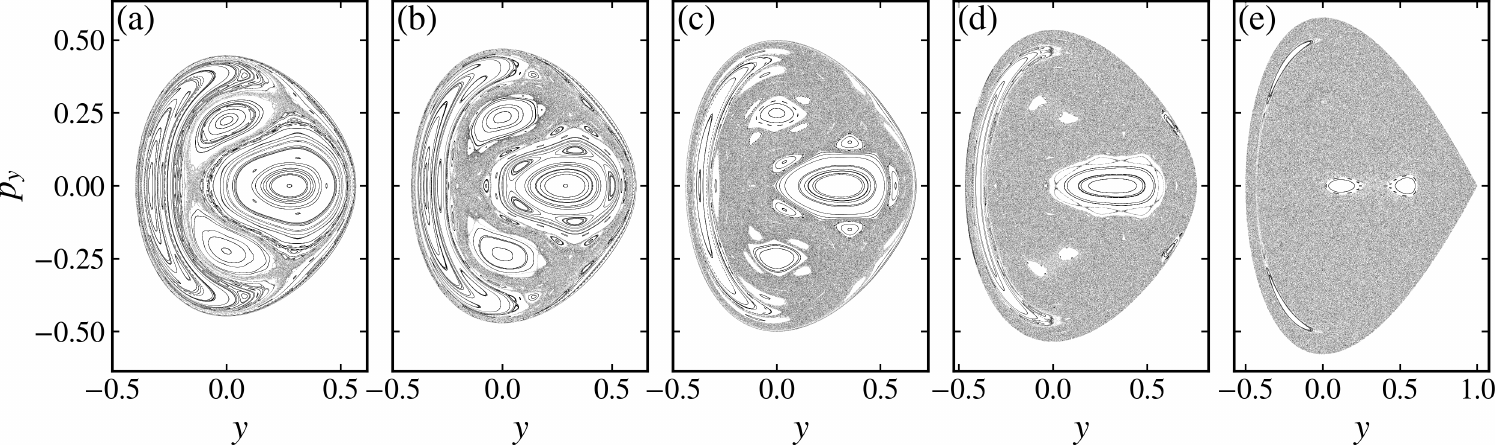}
        \caption{The Poincaré surface of section (PSS) of the Hénon--Heiles system for $x = 0$ and $p_x > 0$ for (a) $E = 1/10$, (b) $E = 1/9$, (c) $E = 1/8$, (d) $E = 1/7$, and (e) $E = 1/6$.}
        \label{fig:PSS}
    \end{figure*}

    The Hénon--Heiles system~\cite{henonheiles} is one of the most paradigmatic models for autonomous, two--degrees--of--freedom Hamiltonian systems. It is a simplified model for the planar motion of a star in an axisymmetric galactic potential, originally introduced to investigate the existence of a third integral of motion. The potential of the Hénon--Heiles system is given by:
    \begin{equation}
        \label{eq:HHpot}
        V(x, y) = \frac{1}{2}(x^2 + y^2) + x^2y - \frac{y^3}{3}.
    \end{equation}
    It can be seen as two harmonic oscillators coupled by a perturbation term $x^2y - y^3/3$. The motion of a test particle with unit mass in the Hénon--Heiles potential is described by the following two--degrees--of--freedom Hamiltonian function:
    \begin{equation}
        \label{eq:Hamiltonian}
        H(p_x, p_y, x, y) = \frac{1}{2}(p_x^2 + p_y^2) + \frac{1}{2}(x^2 + y^2) + x^2y - \frac{y^3}{3} \equiv E,
    \end{equation}
    where $E$ is the conserved total energy, and the equations of motion are
    \begin{equation}
        \label{eq:eqmotion}
        \begin{aligned}
            \dot{x} = p_x,&\quad\dot{y} = p_y,\\
            \dot{p}_x = -x - 2 xy,&\quad\dot{p}_y = - y - x^2 + y^2,
        \end{aligned}
    \end{equation}
    where the dot indicates derivative with respect to time. To solve the system given in Eq.~\eqref{eq:eqmotion}, we need first to find appropriate initial conditions for a given energy level $E$ using Eq.~\eqref{eq:Hamiltonian}. This constraint reduces the dimensionality of the system from four to three. It follows from Eq.~\eqref{eq:Hamiltonian} that
    \begin{equation}
        \label{eq:px}
        p_x(p_y, x, y, E) = \pm\sqrt{2E - p_y^2 -x^2 - y^2 - 2x^2y + \frac{2}{3}y^3}.
    \end{equation}
    One can immediately see that this imposes a constraint on the allowed values of $(y, p_y)$: they need to be chosen in a way that $p_x$ is real. Additionally, Eq.~\eqref{eq:px} gives us two solutions: one positive and one negative. Throughout this paper, we consider the positive solution.

    The Hénon--Heiles potential [Eq.~\eqref{eq:HHpot}] has a finite energy threshold above which the test particles can escape to infinity. For values $E \leq 1/6$, the equipotential curves are closed making the escape impossible. On the other hand, for larger energy values ($E > 1/6$), the equipotential curves are open and the particles can escape to infinity through three exit channels. In this paper, we are not interested in studying the escape of particles in the Hénon--Heiles system, and, therefore, we restrict our analysis to $E < 1/6$. We refer the reader to Refs.~\cite{Blesa2012, Zotos2017, Vallejo2025} and references therein for a survey and recent results on this subject. 

    In order to numerically integrate the equations of motion, we employ the fourth--order Yoshida symplectic integrator~\cite{Yoshida1990}, with the second--order velocity--Verlet integrator~\cite{verlet1967} as the underlying second order step. Throughout this paper, we consider a fixed time step of $\dd{t} = 0.01$ which ensures that the relative energy error, defined by
    \begin{equation}
        \label{eq:Er}
        E_r(t) = \frac{\abs{E(t) - E_0}}{E_0},
    \end{equation}
    remains below $10^{-8}$ and bounded.

    Figure~\ref{fig:PSS} shows the Poincaré surface of section (PSS) for the Hénon--Heiles system defined by $x = 0$ and $p_x > 0$ for different energy values. For low energies, the phase space is dominated by invariant tori and the motion is predominantly regular [Fig.~\ref{fig:PSS}(a)]. As the energy increases, resonant island chains, chaotic layers, and extended chaotic seas become progressively more prominent [Figs.~\ref{fig:PSS}(b)--(d)]. Yet, even when a sizable chaotic component is present, the dynamics is far from uniformly chaotic [Fig.~\ref{fig:PSS}(e)]. Chaotic trajectories may spend very long times trapped near the boundaries of regular islands, displaying intermittent alternation between seemingly regular episodes and irregular bursts. This sticky motion is one of the clearest manifestations of weak chaos in mixed Hamiltonian phase spaces. 
    
    These features make the Hénon--Heiles system an ideal benchmark for testing quantitative indicators of weakly chaotic dynamics for two--degrees--of--freedom Hamiltonian system. The presence of islands in phase space automatically implies a form of non-ergodicity, since trajectories in the chaotic sea cannot enter the islands, and trajectories inside the islands cannot access the chaotic region. Nevertheless, even when considering only the chaotic component of phase space, intermittent trapping near sticky regions effectively makes the dynamics pseudo--ergodic~\cite{ZASLAVSKY2002}. A visual inspection of the Poincaré section reveals the global organization of phase space, but it does not provide a direct characterization of the temporal complexity of individual trajectories, nor does it quantify the intermittent recurrence patterns associated with stickiness. For that, one needs a dynamical measure capable of distinguishing regular motion, strongly chaotic transport, and the long-lived trapping events typical of weak chaos. In Sec.~\ref{sec:rte}, we recall the recurrence-plot construction~\cite{Eckmann1987} and define the recurrence time entropy used in this work~\cite{Little2007,Baptista2010,Kraemer2018}, which is based on the statistics of recurrence times and provides a sensitive tool to characterize weakly chaotic trajectories.

    \section{Recurrence plots and recurrence time entropy}
\label{sec:rte}

    The recurrence plot (RP) was introduced in 1987 by Eckmann \textit{et al.}~\cite{Eckmann1987} as a graphical representation of the recurrences of time series of dynamical systems in its $d$-dimensional phase space. For a given trajectory $\vb{x}_i \in \mathbb{R}^d$ ($i = 1, 2, \ldots, N$) of length $N$, the $N \times N$ recurrence matrix is defined as
    \begin{equation}
        R_{ij} = H\qty(\varepsilon - \|\vb{x}_i - \vb{x}_j\|),
    \end{equation}
    where $H(\cdot)$ is the Heaviside unit step function, $\varepsilon$ is a small threshold and $\|\vb{x}_i - \vb{x}_j\|$ is the distance between states $\vb{x}_i$ and $\vb{x}_j$ in phase space measured in terms of a suitable norm. The most commonly used norms are the Euclidean norm and the maximum (or supremum) norm, defined as
    \begin{equation}
        \begin{aligned}
            \norm{\vb{x}}_2 = \qty(\sum_{i = 1}^d \abs{x_i}^2)^{1/2},\quad\norm{\vb{x}}_\infty = \max_{i}\qty(\abs{x_i}),
        \end{aligned}
    \end{equation}
    respectively. Both of these norms yield similar results. However, the maximum norm is computationally faster and it results in more recurrent points for a fixed threshold $\varepsilon$~\cite{Marwan2007}. Therefore, throughout this paper, we use the maximum norm.

    The recurrent states are represented by the value $1$ in the symmetric, binary recurrence matrix $\vb{R}$, whereas the nonrecurrent ones are represented by the value $0$. Since it is numerically impossible to find \emph{exactly} recurrent states, i.e., $\vb{x}_i = \vb{x}_j$, two states are said to be recurrent if they are sufficiently close to each other up to a distance $\varepsilon$. The distance $\varepsilon$ has to be carefully chosen. If $\varepsilon$ is set too large, nearly every state is recurrent with every other state. On the other hand, if $\varepsilon$ is chosen too small, there will be few recurrent states. Hence, a compromise has to be made between choosing $\varepsilon$ as small as possible while resulting in a sufficient number of recurrent states. There is no general rule on choosing $\varepsilon$. However, a few options have been proposed in the literature and each one of them has its own advantages and disadvantages depending on the purpose of the study. For instance, an alternative is to choose $\varepsilon$ such that the recurrence point density, i.e., the recurrence rate, is fixed~\cite{Zbilut2002, Kraemer2018}. While this eliminates the issue of obtaining few recurrent states, it only shifts the problem to finding the optimal value of the recurrence rate. Another possibility is to define $\varepsilon$ as a fraction of the time series standard deviation~\cite{Thiel2002, Marwan2007, Schinkel2008}. This has been proved efficient when distinguishing between dynamical regimes and analyzing dynamical transition~\cite{Ngamga2007, Ngamga2008, Sales2023, Souza2024}. Therefore, unless otherwise stated, we set the threshold to be a small percentage of the time series standard deviation: $\varepsilon = 0.05\sigma$.

    Graphically, recurrent states are represented by colored dots, and the recurrence matrix displays characteristic patterns that depend on the nature of the underlying dynamics. These patterns are mainly organized into four basic structures: (i) isolated recurrence points, (ii) diagonal lines, (iii) vertical lines, and (iv) white vertical gaps corresponding to non-recurrent states. Several quantitative measures have been introduced based on these structures, and together they constitute what is known as recurrence quantification analysis (RQA). For detailed discussions of RQA and its applications, we refer the reader to Refs.~\cite{Marwan2007, Marwan2008, Marwan2025, Marwan2026} and the references therein. To construct the recurrence matrix, instead of using the full trajectory $(x(t), y(t), p_x(t), p_y(t))$ in phase space, we restrict the dynamics to the PSS defined by $x = 0$ and $p_x > 0$, and then consider the corresponding sequence of intersection points $(y_n, p_{y,n})$. That is, the recurrence analysis is performed on the two--dimensional sequence generated by successive crossings of the chosen surface of section. We emphasize that no phase space reconstruction is involved. Delay embedding is required when only a scalar observable is measured and the phase space has to be reconstructed from it~\cite{Marwan2007}. In our case the state of the system is known at every crossing, so that the recurrence matrix is computed directly from the sequence of state vectors. This is also advantageous, since embedding may introduce spurious correlations that are reflected in the RP and affect the quantification of its structures~\cite{Marwan2007}.

    \begin{figure*}[t!]
        \centering
        \includegraphics[width=\linewidth]{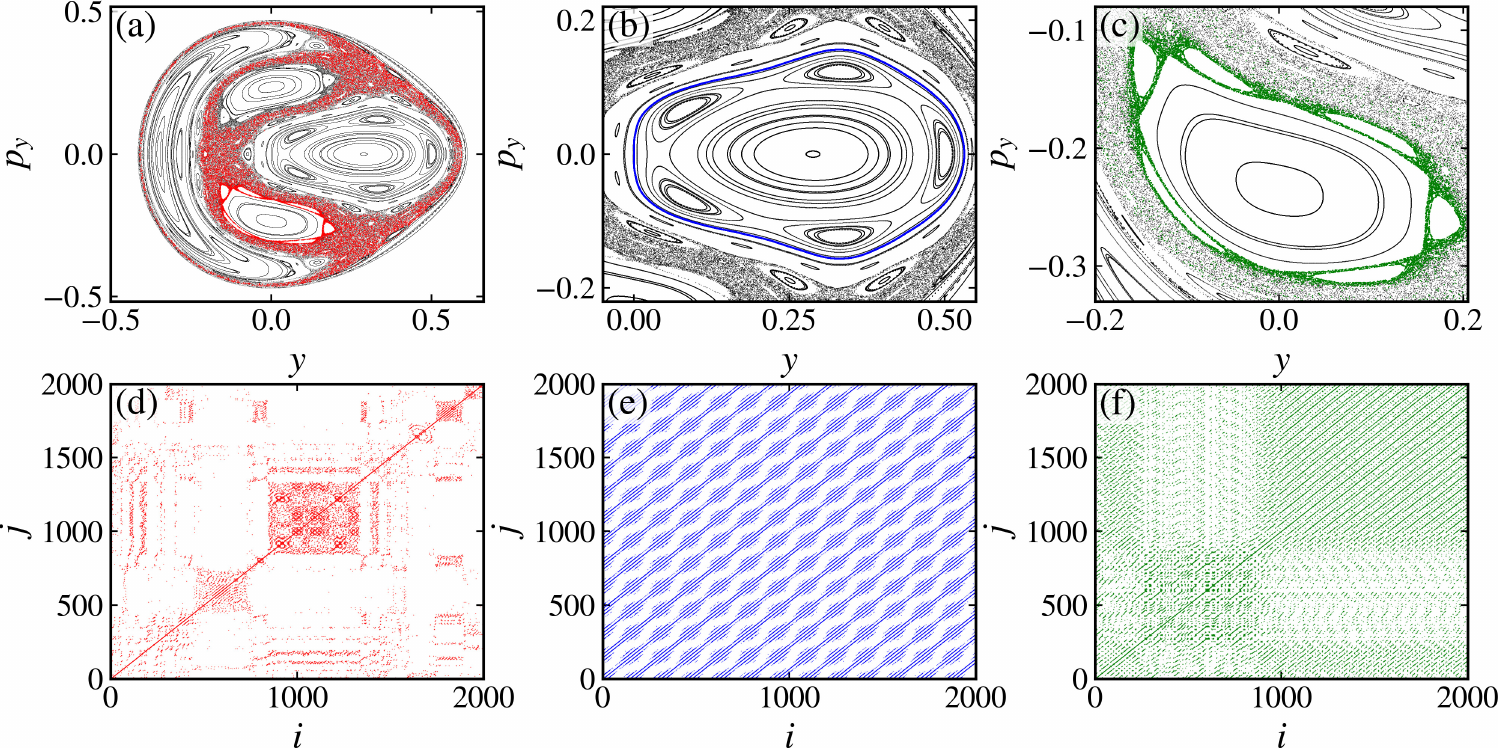}
        \caption{Example of a (a) chaotic (in red), (b) quasi-periodic (in blue), and (c) sticky trajectory (in green). (d)--(f) The respective recurrence matrices. The energy level is $E = 1/9$ and the initial conditions are (red) $(x, y, p_x, p_y) = (0.0, -0.15, p_x, 0.0)$, (blue) $(x, y, p_x, p_y) = (0.0, 0.0, p_x, 0.0)$, and (green) $(x, y, p_x, p_y) = (0.0, -0.125, p_x, -0.127)$, with $p_x$ calculated according to Eq.~\eqref{eq:px}.}
        \label{fig:recmats}
    \end{figure*}

    Figure~\ref{fig:recmats}(a) shows three representative trajectories of the Hénon--Heiles system at $E = 1/9$, while Figs.~\ref{fig:recmats}(b)--\ref{fig:recmats}(d) display the corresponding recurrence matrices obtained from the first 2000 crossings of the chosen PSS. The blue trajectory in Fig.~\ref{fig:recmats}(a) is quasi-periodic, and its recurrence matrix [Fig.~\ref{fig:recmats}(e)] is dominated by long, nearly uninterrupted diagonal lines, as expected for highly regular motion. By contrast, the red trajectory is chaotic, and its recurrence matrix [Fig.~\ref{fig:recmats}(d)] is characterized by much shorter diagonal segments, reflecting the stronger irregularity of the underlying dynamics. The intermediate case is the sticky trajectory, shown in green in Fig.~\ref{fig:recmats}(a). In this case, the orbit remains temporarily trapped near the regular structures located between the lower period-1 island and its period-6 satellite islands before exploring a wider chaotic region. Accordingly, its recurrence matrix [Fig.~\ref{fig:recmats}(f)] exhibits diagonal lines of intermediate length: longer than those of the chaotic trajectory, but shorter and less persistent than those of the quasi-periodic one. These examples illustrate that different dynamical regimes generate distinct recurrence patterns, which motivates the use of RQA quantifiers for their characterization.
    
    Among the many measures available in RQA, we are particularly interested in those that are sensitive to the temporal organization of recurrence times. This is essential in mixed Hamiltonian systems, where weak chaos is typically associated with intermittency and stickiness, leading to broad and highly nonuniform distributions of recurrence times~\cite{Karney1983,Cristadoro2008,Abud2013,Lozej2020}. For this reason, we use the recurrence time entropy (RTE), computed from the distribution of white vertical lines in the RP. The white vertical lines in an RP, i.e., the vertical gaps between two diagonal lines, provide an estimate of the recurrence times of a trajectory~\cite{Zou2007b, Zou2007, Baptista2010,Ngamga2012}. Therefore, the RTE is defined as the Shannon entropy of the distribution of white vertical lines:
    \begin{equation}
        \RTE = -\sum_{\ell=\ell_{\mathrm{min}}}^{\ell_{\mathrm{max}}} p(\ell) \log p(\ell),
    \end{equation}
    where $\ell_{\text{min}}$ ($\ell_{\text{max}}$) is the length of the shortest (longest) white vertical line, $p(\ell) = P(\ell) / \mathcal{N}$ is the relative distribution of white vertical lines and $\mathcal{N}$ is the total number of white vertical line segments. Formally, $P(\ell)$ is defined as
    \begin{equation}
        \label{eq:histogram}
        P(\ell) = \sum_{i,j=1}^NR_{i, j-1}R_{i,j+\ell}\prod_{k=0}^{\ell - 1}\qty(1 - R_{i, j+k}).
    \end{equation}
    The evaluation of the histogram given by Eq.~\eqref{eq:histogram} should be done carefully. Due to the finite size of an RP, the distribution of white vertical lines might be biased by the border lines, i.e., the lines that begin and end at the borders of an RP. These lines are cut short by the borders of the RP and their length is measured incorrectly. This influences measures such as the RTE~\cite{Kraemer2019}. To avoid such border effects, we exclude from the distribution the white vertical lines that begin and end at the border of the RP.

    The choice of the RTE among the RQA quantifiers is supported by previous comparisons. For the standard map, the Pearson correlation coefficient between the largest Lyapunov exponent and eleven RQA quantifiers has been computed from time series of length $N = 5000$ in Ref.~\cite{SalesThesis2023}, and the determinism and the RTE yielded the largest correlations in absolute value, $-0.96$ and $0.95$, respectively. The RTE has also been employed together with the recurrence rate and the determinism to characterize transitions between dynamical regimes~\cite{Gabrick2023}, and together with the Hurst exponent, a quantifier that is not derived from RPs, to characterize weak chaos in fractional maps, where both assume lower values in the weakly chaotic regions~\cite{Borin2025}.

    \begin{figure*}[t!]
        \centering
        \includegraphics[width=\linewidth]{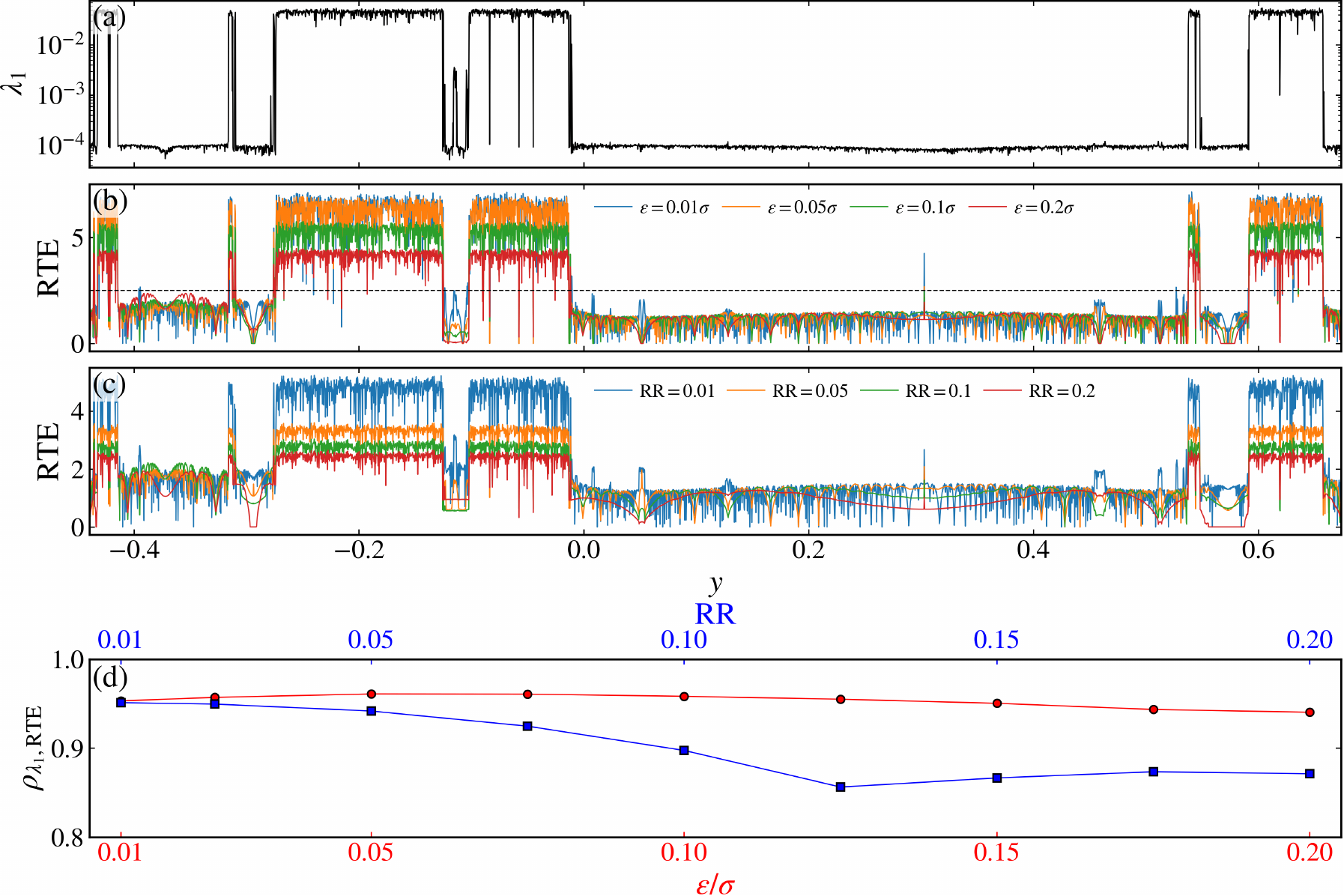}
        \caption{(a) The largest Lyapunov exponent for $t_{\mathrm{total}} = 10^5$, and the RTE for $N_{\mathrm{cross}} = 10^4$ with the threshold determined (b) as a fraction of the time series' standard deviation, $\sigma$, and (c) by imposing a fixed recurrence rate, RR, as a function of $y$ with $x = p_y = 0$ and $p_x = p_x(p_y, x, y , E)$ [Eq.~\eqref{eq:px}] for $E = 1/8$. The horizontal dashed black line in panel (b) corresponds to the value of $\RTE = 2.5$, which will be used as the threshold for chaos detection in later sections. (d) The Pearson correlation coefficient between $\lambda_1$ and the RTE for the two methods of determining $\varepsilon$. Each method has its own abscissa: the circles (red) are read on the bottom axis, which gives $\varepsilon$ in units of $\sigma$, and the squares (blue) are read on the top axis, which gives the imposed recurrence rate. The two quantities are distinct parameters that happen to span the same numerical range, and the abscissa must not be read as a common scale.}
        \label{fig:LLERTE_vs_y}
    \end{figure*}

    \section{Characterization of weakly chaotic dynamics}
    \label{sec:hh_results}

    \begin{figure*}[t!]
        \centering
        \includegraphics[width=\linewidth]{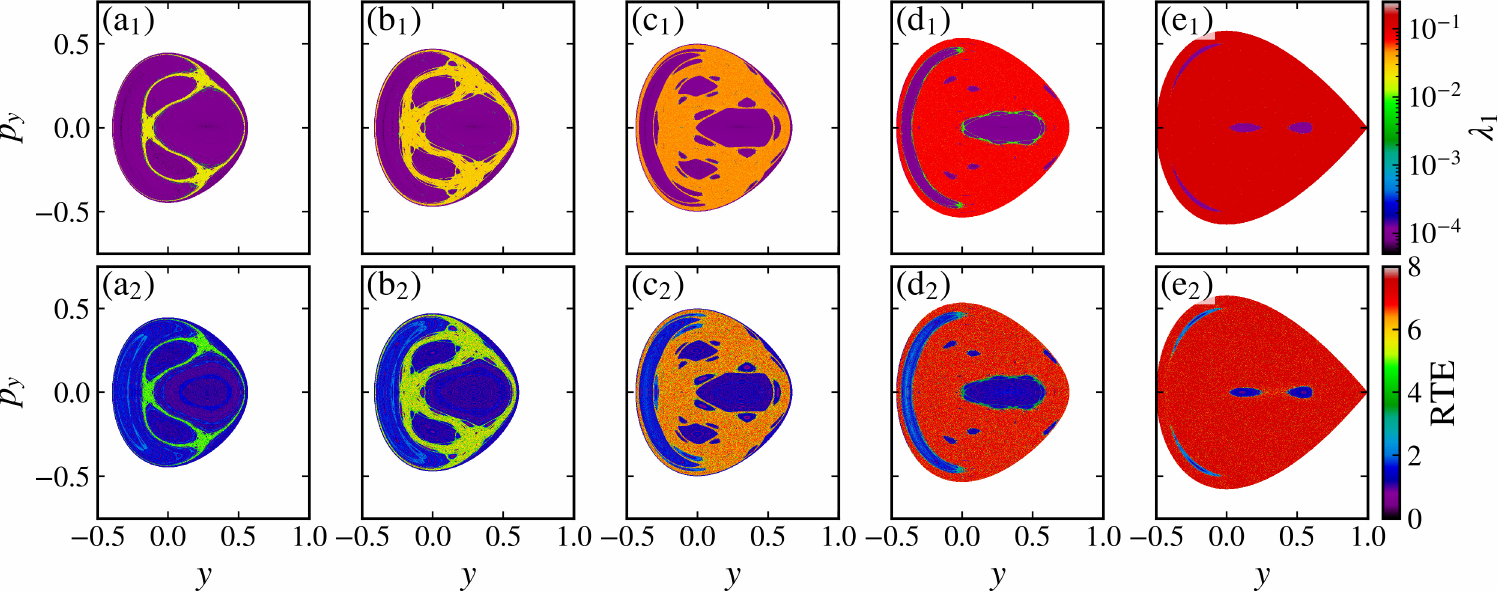}
        \caption{(Top row) The largest Lyapunov exponent for $t_{\mathrm{total}} = 10^5$ and (bottom row) the RTE for $N_{\mathrm{cross}} = 10^4$ calculated in a grid of initial conditions uniformly distributed in $(y, p_y)\in[-0.5, 1.0]\times[-0.75, 0.75]$ with $x = 0$ and $p_x = p_x(p_y, x, y, E)$ [Eq.~\eqref{eq:px}] for (a) $E = 1/10$, (b) $E = 1/9$, (c) $E = 1/8$, (d) $E = 1/7$, and (e) $E = 1/6$.}
            \label{fig:gridLLERTE}
    \end{figure*}

    In this section, we apply the RTE to the PSS of the Hénon--Heiles Hamiltonian and show that it distinguishes regular from chaotic motion while also capturing weakly chaotic dynamics. As a benchmark, we compare it with the largest Lyapunov exponent (LLE). 
    
    Figure~\ref{fig:LLERTE_vs_y}(a) shows the LLE, and Figs.~\ref{fig:LLERTE_vs_y}(b) and \ref{fig:LLERTE_vs_y}(c) show the RTE, both computed as functions of $y$ for fixed $x = 0$, $p_y = 0$, and $E = 1/8$. For the LLE we consider a total integration time of $t_{\mathrm{total}} = 10^5$ and $N_{\mathrm{cross}} = 10^4$ for the RTE. This value of $N_{\mathrm{cross}}$ is used throughout the paper, and the dependence of the RTE on the number of crossings is examined in Appendix~\ref{sec:appendix}. We also compute the RTE for different values of the recurrence threshold $\varepsilon$, determined in the two ways discussed above: as a fraction of the time series' standard deviation [Fig.~\ref{fig:LLERTE_vs_y}(b)] and by imposing a fixed recurrence rate [Fig.~\ref{fig:LLERTE_vs_y}(c)]. The results show a clear positive correlation between the two quantities. In regions where $\lambda_1 > 0$, the RTE takes relatively large values, consistent with chaotic dynamics. By contrast, within regular regions, where $\lambda_1$ is close to zero, the RTE remains small. Although changing the threshold modifies the numerical values of the RTE, especially in the chaotic regions, it does not alter its qualitative dependence on $y$, and this holds for both methods. This indicates that the RTE is robust with respect to the choice of threshold and captures the main dynamical structures of the system in a way consistent with the LLE. This agreement is quantified in Fig.~\ref{fig:LLERTE_vs_y}(d), where we show the Pearson correlation coefficient between the LLE and the RTE obtained for different threshold values and for both methods. Since the two methods are parametrized by different quantities, each of them is read on its own abscissa in this panel: the circles correspond to the threshold set as a fraction of the standard deviation and are read on the bottom axis, which gives $\varepsilon/\sigma$, whereas the squares correspond to the imposed recurrence rate and are read on the top axis, which gives RR. The correlation is high in all cases: it always exceeds $0.94$ when $\varepsilon$ is set as a fraction of the standard deviation and $0.85$ when it is set by fixing the recurrence rate, which confirms that the two measures display very similar variations along the scanned initial conditions regardless of how the threshold is chosen. The largest correlation is found for $\varepsilon = 0.05\sigma$, which is the value adopted throughout this paper, indicating that this choice of threshold yields the closest correspondence between the recurrence--time statistics and the instability measured by the LLE. We return to the comparison between the two methods in Sec.~\ref{sec:threedof}, for a system whose surface of section is four--dimensional and where the choice of the threshold plays a more visible role.

   To further examine the correspondence between the LLE and the RTE, we compute both quantities on a grid of uniformly distributed initial conditions in $(y,p_y)\in[-0.5,1.0]\times[-0.75,0.75]$, with $x=0$ and $p_x=p_x(p_y,x,y,E)$, for five energy values and $t_{\mathrm{total}} = 10^5$ and $N_{\mathrm{cross}} = 10^4$, respectively. The results are shown in Fig.~\ref{fig:gridLLERTE}, where the top row displays the LLE and the bottom row the RTE. The two measures identify the same main phase space structures. The RTE, however, exhibits a higher variation in the chaotic region, due to the RTE being more sensitive to the trapping regions. Regular islands are associated with small values of both the LLE and the RTE, whereas the chaotic sea is characterized by larger values. As the energy increases, the regular regions shrink and the chaotic component expands, and this trend is captured consistently by both indicators. Near the borders of the regular islands, the RTE typically assumes intermediate values, reflecting the weakly chaotic dynamics that arises in these sticky transition regions. We use the term sticky transition layer to designate the set of initial conditions whose RTE takes values intermediate between those attained inside the regular islands and those of the chaotic sea. These layers are directly visible in Fig.~\ref{fig:gridLLERTE}(d$_2$), where the chaotic sea appears in red and the regular islands in dark blue, and a narrow green band, corresponding to intermediate values of the RTE, traces the boundary of every island. The threshold $\RTE = 2.5$, adopted below to classify trajectories as regular or chaotic, lies above the values found inside the islands and below those of the transition layers, so that the latter are classified as chaotic, as they should be, since sticky orbits are chaotic orbits temporarily trapped near regular structures. This global agreement supports the use of the RTE as a recurrence--based indicator of the mixed phase space structure in the Hénon--Heiles system.

\begin{figure*}[t!]
        \centering
        \includegraphics[width=\linewidth]{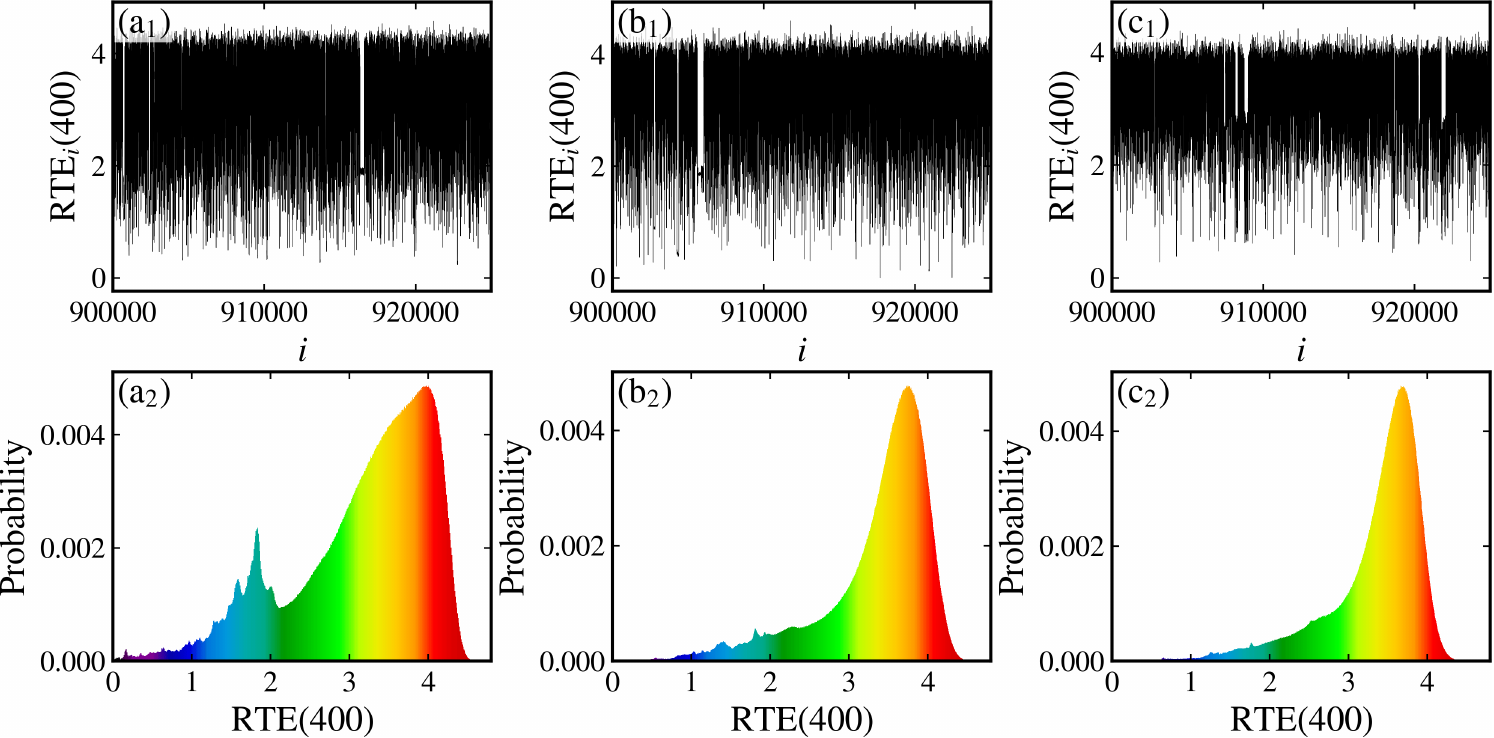}
        \caption{(Top row) The finite time RTE time series and (bottom row) the probability distribution of their values for $x = 0$, $y = -0.15$, $p_y = 0$ and $p_x = p_x(p_y, x, y, E)$ [Eq.~\eqref{eq:px}] with (a) $E = 1/9$, (b) $E = 1/8$, and (c) $E = 1/7$.}
        \label{fig:ftrte}
    \end{figure*}

    We now analyze the temporal fluctuations of the RTE along a single chaotic trajectory. For this purpose, we compute the finite–time RTE in consecutive windows of $400$ crossings of the PSS, obtaining a time series $\RTE(400)$ for each energy. The total length of each trajectory is $10^9$ crossings. The results are shown in the top row of Fig.~\ref{fig:ftrte}. For all three energies, the finite--time RTE displays strong fluctuations, with abrupt drops from large values to smaller ones. These drops indicate time intervals in which the recurrence--time statistics become less complex, as expected when the chaotic trajectory is temporarily trapped near regular islands or close to their hierarchical structures. Therefore, the finite--time RTE does not only separate regular and chaotic regions globally, as in Fig.~\ref{fig:gridLLERTE}, but also resolves the intermittent changes in the local character of a single chaotic trajectory.

    The probability distributions shown in the bottom row of Fig.~\ref{fig:ftrte} make this intermittency clearer. For $E=1/9$, the distribution is broad and displays several local maxima, especially at smaller and intermediate values of $\RTE(400)$. This indicates that the trajectory alternates between intervals of extended exploration of the chaotic component and intervals of temporary trapping near regular structures, producing distinct recurrence--time statistics over finite windows. As the energy increases to $E=1/8$ and $E=1/7$, the distribution becomes increasingly dominated by large values of $\RTE(400)$, while the relative weight of the low--RTE events decreases. This is consistent with the enlargement of the chaotic component observed in Fig.~\ref{fig:gridLLERTE}: the trajectory spends a larger fraction of time in regions with high recurrence--time complexity, but still exhibits intermittent visits to sticky regions, visible as the low--RTE tail of the distributions.

    The important point is that the finite--time RTE converts the qualitative notion of stickiness into a measurable time-dependent signal. Large values of $\RTE(400)$ correspond to intervals in which the orbit explores the chaotic sea and the recurrence times are broadly distributed. Smaller values correspond to intervals in which the orbit is temporarily trapped near regular structures, so that the recurrence--time distribution becomes less complex. The coexistence of these regimes produces the asymmetric and, in some cases, multimodal distributions observed in Fig.~\ref{fig:ftrte}. This behavior is analogous to what has been reported for two--dimensional area--preserving maps, where finite--time RTE distributions become multimodal in the presence of sticky motion and the different peaks are associated with different levels of the islands-around-islands hierarchy~\cite{Sales2023, Souza2024, Viana2025}. In the next step, we use this finite--time signal to identify the phase space regions responsible for the low-- and high--RTE episodes and to quantify the corresponding trapping--time statistics.

\begin{figure*}[t!]
        \centering
        \includegraphics[width=\linewidth]{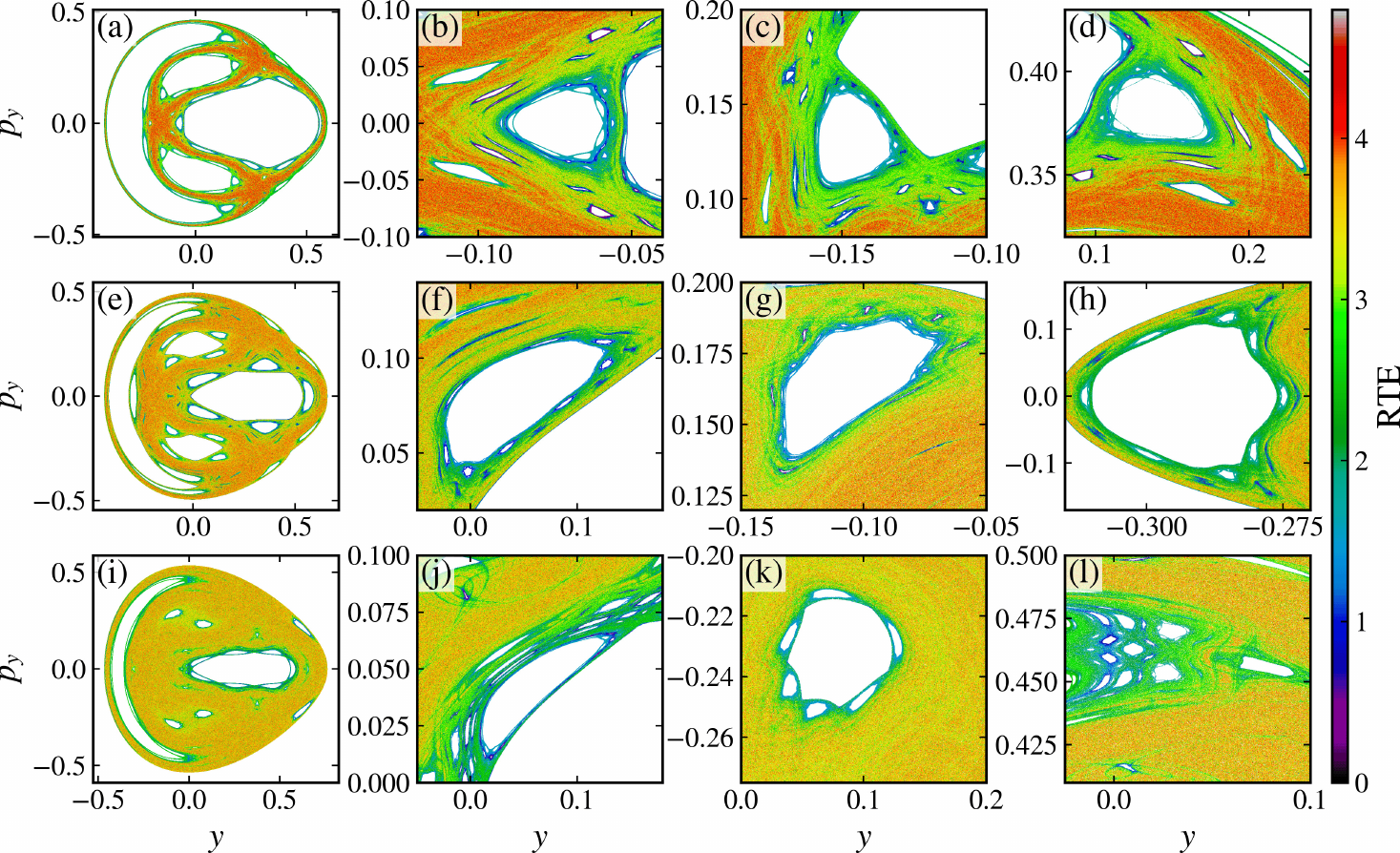}
        \caption{PSS points of the chaotic trajectory used in Fig.~\ref{fig:ftrte}, colored according to the finite--time value of $\RTE(400)$. Panels (a)--(d), (e)--(h), and (i)--(l) correspond to $E = 1/9$, $E = 1/8$, and $E = 1/7$, respectively. For each energy, the first panel shows a global view of the explored chaotic component, while the remaining panels show magnifications of regions near regular islands and their surrounding sticky layers.}
        \label{fig:ftrte_pss}
    \end{figure*}

    To identify where the different values of the finite--time RTE are generated in phase space, Fig.~\ref{fig:ftrte_pss} shows the PSS points associated with the same trajectory used in Fig.~\ref{fig:ftrte}, colored according to the value of $\RTE(400)$ computed in each finite--time window. For each energy, the first panel gives a global view of the chaotic component, while the following panels show magnifications of regions surrounding regular islands. This representation connects the temporal fluctuations of Fig.~\ref{fig:ftrte} with the corresponding phase space structures. For $E=1/9$ [Figs.~\ref{fig:ftrte_pss}(a)--(d)], the chaotic trajectory visits a broad chaotic layer but also spends long intervals near the boundaries of regular islands. The regions close to the islands are predominantly associated with smaller and intermediate values of $\RTE(400)$, whereas larger values occur when the trajectory explores the surrounding chaotic sea. The magnifications show that these low-- and intermediate--RTE values are not randomly distributed: they are organized around the island boundaries and along thin structures associated with the sticky region. This explains the broad and structured distribution observed in Fig.~\ref{fig:ftrte}(a$_2$).

    The same mechanism persists for $E=1/8$ [Figs.~\ref{fig:ftrte_pss}(e)--(h)]. The chaotic component is larger, but the finite--time RTE still decreases near regular islands. Therefore, even though the trajectory is chaotic, its recurrence--time statistics are strongly affected by temporary trapping near regular structures. The magnified panels show layers of different RTE values around the islands, indicating that the finite--time RTE is sensitive not only to whether the orbit is in the chaotic sea or near an island, but also to different recurrence regimes within the sticky layer. For $E=1/7$ [Figs.~\ref{fig:ftrte_pss}(i)--(l)], the chaotic component occupies an even larger portion of the accessible PSS, and high values of $\RTE(400)$ dominate most of the explored region. Nevertheless, low-- and intermediate--RTE episodes remain localized near regular structures and their surrounding sticky layers. This confirms that the low--RTE tail in Fig.~\ref{fig:ftrte}(c$_2$) is not a statistical artifact of the finite--time computation, but is generated by specific regions of the mixed phase space.

    Overall, Fig.~\ref{fig:ftrte_pss} shows that the finite--time RTE provides a phase space-resolved description of stickiness. Large values of $\RTE(400)$ are associated with extended exploration of the chaotic component, while smaller values are concentrated near regular islands, where the trajectory undergoes temporary trapping. Thus, the fluctuations and asymmetric distributions shown in Fig.~\ref{fig:ftrte} have a direct geometrical origin in the mixed structure of the PSS.

    Having identified the phase space origin of the low-- and high--RTE episodes, we now quantify how long the trajectory stays in each regime. We define low-- and high--RTE episodes directly from the finite--time RTE time series: consecutive windows with $\RTE(400) < 2$ are associated with trapping near regular structures, whereas consecutive windows with $\RTE(400) > 3$ correspond to intervals of extended exploration of the chaotic component. These two values are read off the distributions shown in the bottom row of Fig.~\ref{fig:ftrte}: for $E = 1/9$ the distribution is bimodal, with a secondary maximum near $\RTE(400) \approx 1.8$, associated with trapping, and a main maximum near $\RTE(400) \approx 3.9$, associated with chaotic exploration, separated by a minimum in between. The intermediate band $2 \leq \RTE(400) \leq 3$ is deliberately left out of the classification, so that only unambiguous episodes are counted. To avoid counting short-lived fluctuations, we only consider an episode when at least three consecutive windows satisfy the corresponding condition. For each case, we measure the duration $\tau$ of these episodes and compute the cumulative distribution
    \begin{equation}
        Q(\tau) = \frac{N_{\tau}}{N_t},
    \end{equation}
    where $N_{\tau}$ is the number of episodes with duration larger than $\tau$ and $N_t$ is the total number of episodes.

\begin{figure*}[t!]
        \centering
        \includegraphics[width=\linewidth]{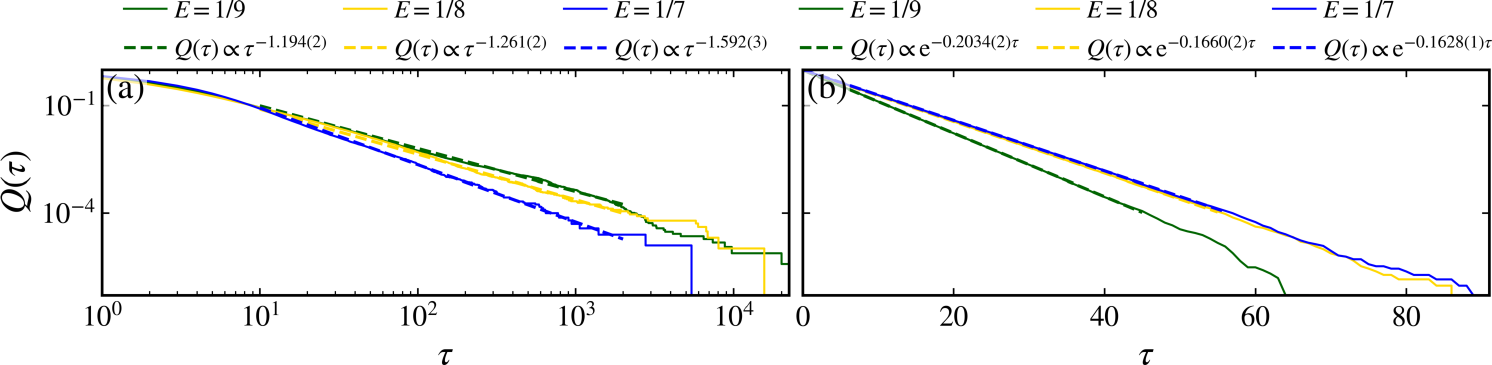}
        \caption{Cumulative distribution of episode durations for (a) $\RTE(400) < 2$ and (b) $\RTE(400) > 3$. In panel (a), the low--RTE episodes are associated with trapping near regular structures and $Q(\tau)$ decays algebraically as $Q(\tau) \propto \tau^{-\gamma}$, with ($E = 1/9$) $\gamma = 1.194(2)$, ($E = 1/8$) $\gamma = 1.261(2)$, and ($E = 1/7$) $\gamma = 1.592(3)$. In panel (b), the high--RTE episodes are associated with extended exploration of the chaotic component and $Q(\tau)$ decays exponentially as $Q(\tau) \propto \mathrm{e}^{-\kappa \tau}$, with ($E = 1/9$) $\kappa = 0.2034(2)$, ($E = 1/8$) $\kappa = 0.1660(2)$, and ($E = 1/7$) $\kappa = 0.1628(1)$.}
        \label{fig:traptimes}
    \end{figure*}

        Figure~\ref{fig:traptimes} shows $Q(\tau)$ for the two regimes. For $\RTE(400) < 2$ [Fig.~\ref{fig:traptimes}(a)], the cumulative distributions display algebraic tails,
    \begin{equation}
        \label{eq:gamma}
        Q(\tau) \propto \tau^{-\gamma},
    \end{equation}
    with $\gamma = 1.194(2)$ for $E=1/9$, $\gamma = 1.261(2)$ for $E=1/8$, and $\gamma = 1.592(3)$ for $E=1/7$. This power-law decay is the expected statistical signature of stickiness in mixed Hamiltonian systems: trapping events near regular structures do not have a characteristic time scale, and long episodes occur with non-negligible probability. By contrast, for $\RTE(400) > 3$ [Fig.~\ref{fig:traptimes}(b)], the cumulative distributions decay exponentially,
    \begin{equation}
        \label{eq:kappa}
        Q(\tau) \propto \mathrm{e}^{-\kappa \tau},
    \end{equation}
    with $\kappa = 0.2034(2)$ for $E=1/9$, $\kappa = 0.1660(2)$ for $E=1/8$, and $\kappa = 0.1628(1)$ for $E=1/7$. Therefore, the high--RTE episodes are characterized by a typical time scale, as expected for intervals in which the trajectory explores the more strongly chaotic part of the accessible component.

The contrast between the algebraic decay for $\RTE(400) < 2$ and the exponential decay for $\RTE(400) > 3$ confirms that the finite--time RTE separates two dynamically distinct regimes, and this dichotomy has a direct counterpart in the theory of transient chaos. For weakly chaotic systems, the invariant chaotic saddle can be split into a hyperbolic component, whose escape and recurrence statistics decay exponentially, and a nonhyperbolic component, formed by the neighborhoods of the regular islands, which is responsible for the asymptotic power--law decay~\cite{Altmann2008, Altmann2009}. The two regimes identified by the finite--time RTE correspond to these two components. The high--RTE episodes are intervals in which the trajectory evolves in the hyperbolic part of the chaotic component, where the motion decorrelates rapidly and the durations are exponentially distributed, whereas the low--RTE episodes are generated by trapping in the nonhyperbolic part, where partial transport barriers (cantori) and hierarchical island chains constrain the motion and produce heavy--tailed statistics~\cite{Karney1983, Mackay1984, Meiss1986}. In this interpretation, $\kappa$ plays the role of an escape rate associated with the hyperbolic component. Thus, the finite--time RTE not only identifies where stickiness occurs in phase space, but also recovers its characteristic trapping--time statistics.

    Three elements support the identification of these episodes with stickiness. First, the low--RTE episodes are not distributed uniformly over the chaotic component: when projected back onto the PSS, they concentrate around the regular islands and their surrounding hierarchical structures (Fig.~\ref{fig:ftrte_pss}). Second, their durations follow the algebraic decay expected for trapping in a mixed phase space, whereas the durations of the high--RTE episodes decay exponentially (Fig.~\ref{fig:traptimes}). Third, the multimodal structure of the finite--time RTE distributions (Fig.~\ref{fig:ftrte}) are similar to the multimodal distributions of finite--time Lyapunov exponents reported for systems with a hierarchy of islands~\cite{Szezech2005, Harle2007}, and the correspondence between the two has been established for the standard map~\cite{Sales2023}, where the modes of the finite--time RTE distribution were associated with different hierarchical levels of the islands--around--islands structure.
    
\begin{figure*}[t!]
        \centering
        \includegraphics[width=\linewidth]{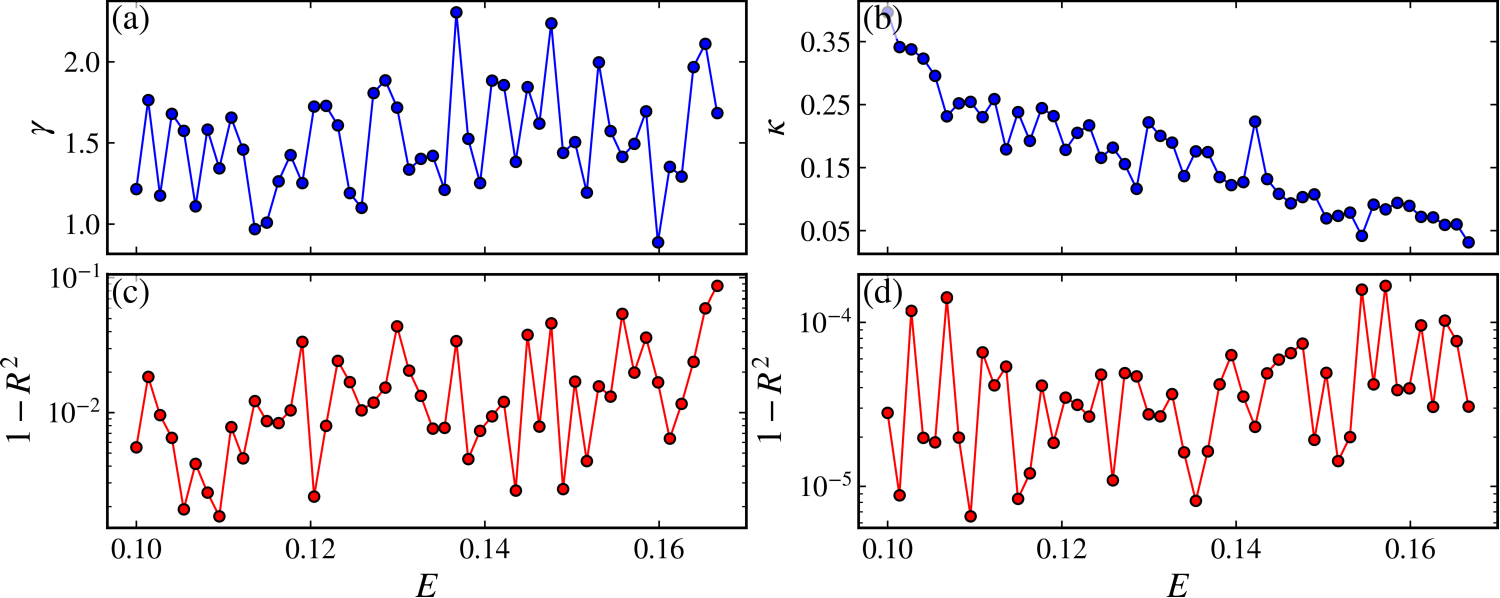}
        \caption{(a) The exponent $\gamma$ [Eq.~\eqref{eq:gamma}] and (b) the exponent $\kappa$ [Eq.~\eqref{eq:kappa}] as functions of the energy $E$. For each energy value, finite--time RTE time series were computed as in Fig.~\ref{fig:ftrte}, and cumulative distributions of low-- ($\RTE < 2$) and high--RTE ($\RTE > 3$) episodes were constructed as in Fig.~\ref{fig:traptimes}. The corresponding exponents were extracted from fits to these distributions. Panels (c) and (d) show $1 - R^2$ for the fitted curves, where smaller values indicate better agreement between the fit and the data.}
        \label{fig:exponents}
    \end{figure*}

    It is instructive to compare the values of $\gamma$ obtained here with those reported for area--preserving maps, for which the trapping statistics have been studied in detail. The Markov tree model of the islands--around--islands hierarchy predicts $\gamma = 1.96$~\cite{Meiss1986}, renormalization arguments at the breakup of the golden mean torus predict $\gamma = 3$, and the scaling properties of self--similar island chains relate $\gamma$ to the geometry of the hierarchy~\cite{ZASLAVSKY2002}. These mechanisms coexist in generic systems, and finite--time numerical estimates of $\gamma$ are typically found in the interval $1.5 \lesssim \gamma \lesssim 2.5$~\cite{Altmann2006}. Furthermore, an asymptotically universal exponent close to $1.57$ has been conjectured for two--degrees--of--freedom Hamiltonian systems, the system--dependent values obtained in finite--time numerical experiments being attributed to the slow convergence towards this asymptotic regime~\cite{Cristadoro2008}. The exponents obtained here are compatible with this scenario: for the three energies considered above $\gamma$ lies between $1.19$ and $1.59$, and over the broader energy range examined below it varies between approximately $0.9$ and $2.3$, with most values lying between $1$ and $2$ (Fig.~\ref{fig:exponents}). We stress, however, that the durations measured here are those of the low--RTE episodes, defined through the finite--time RTE, and not Poincaré recurrence times to a fixed region of the PSS. The two quantities are closely related, since both measure how long a trajectory remains in the neighborhood of the regular structures, but they are not identical, and the recurrence time statistics are known to depend on the choice and on the size of the recurrence region~\cite{Altmann2004, Altmann2005b}. The comparison above should therefore be understood at the level of the range of the exponents and of their qualitative behavior, and not as an identification of the two quantities.

    To investigate how the statistical properties of these episodes change with the energy, we repeat the same procedure described above for a range of values $E \in [1/10,1/6]$ and extract the corresponding exponents $\gamma$ and $\kappa$ from fits to the cumulative distributions. The results are shown in Fig.~\ref{fig:exponents}. Panel (a) displays the exponent $\gamma$ associated with the algebraic decay of low--RTE episodes [$\RTE(400)<2$], while panel (b) shows the exponent $\kappa$ associated with the exponential decay of high--RTE episodes [$\RTE(400)>3$]. Panels (c) and (d) display the corresponding values of $1-R^2$ for the fits, where smaller values indicate better agreement between the fitted function and the data.

    The behavior of $\gamma$ in Fig.~\ref{fig:exponents}(a) is strongly nonmonotonic and exhibits pronounced oscillations as the energy varies. This behavior is expected in mixed Hamiltonian systems because the phase space structure changes nontrivially with the energy. As $E$ increases, regular islands deform, resonances appear and disappear, chaotic layers merge, and the hierarchy of sticky structures surrounding the islands is reorganized. Since the exponent $\gamma$ quantifies the long-tailed statistics of trapping near these structures, even relatively small geometrical changes in phase space may significantly alter the measured value of the algebraic decay. Furthermore, the estimation of $\gamma$ is intrinsically sensitive to the longest trapping events, which are comparatively rare and dominate the tail of the distribution. Consequently, the power-law fitting is naturally less stable than the exponential case, as reflected by the larger fluctuations in $1-R^2$ shown in Fig.~\ref{fig:exponents}(c), especially for energies where the sticky dynamics generates very long tails. This behavior is also consistent with the system dependence of the finite--time estimates of $\gamma$ reported for area--preserving maps~\cite{Altmann2006, Cristadoro2008}. Here, the energy plays the role of a continuous control parameter, so that this dependence can be followed as the hierarchical organization of the phase space is progressively modified.
    
\begin{figure}[t!]
        \centering
        \includegraphics[width=\linewidth]{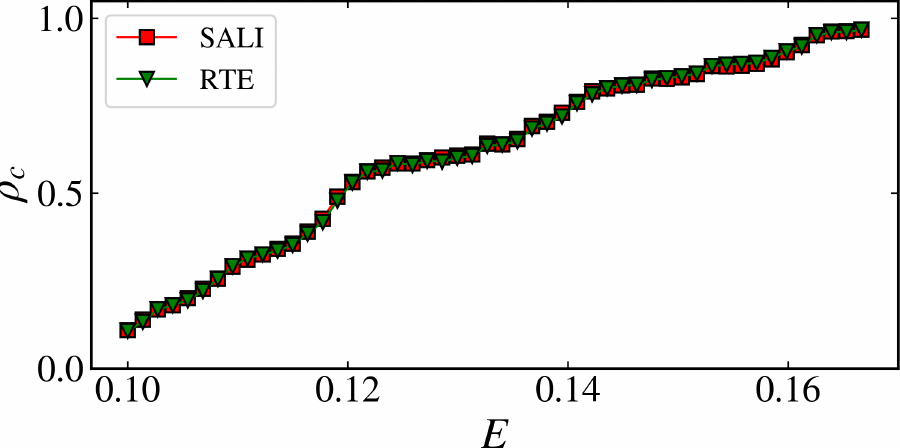}
        \caption{Proportion of chaotic orbits in an ensemble of $10^4$ randomly selected initial conditions as a function of the energy, computed using the (green triangles) RTE and (red squares) SALI as chaos indicators. For SALI, each initial condition was evolved up to $t_f = 10^4$ and classified as chaotic when $\mathrm{SALI}(t_f) < 10^{-10}$. For the RTE, $n_c = 10^4$ crossings of the Poincaré surface of section defined by $x = 0$ and $p_x > 0$ were computed for each initial condition, and trajectories with $\RTE(n_c) > 2.5$ were classified as chaotic [see Fig.~\ref{fig:LLERTE_vs_y}(b)].}
        \label{fig:rhoc}
    \end{figure}

    By contrast, the behavior of $\kappa$ in Fig.~\ref{fig:exponents}(b) is considerably smoother and displays an overall decreasing trend with increasing energy. Since $\kappa$ acts as an escape rate from the strongly chaotic regime, it defines a characteristic time scale
    \begin{equation*}
        T_c \sim \frac{1}{\kappa},
    \end{equation*}
    that measures the typical duration of high--RTE episodes. Therefore, the decrease of $\kappa$ with energy implies that chaotic trajectories tend to remain for longer intervals in the strongly chaotic component as $E$ increases. This behavior is consistent with the global phase space evolution observed in Fig.~\ref{fig:gridLLERTE}, where the chaotic component expands progressively with energy, and with Fig.~\ref{fig:rhoc}, where the proportion of chaotic trajectories increases with $E$. In other words, as the chaotic sea grows, trajectories not only become more likely to be chaotic, but also tend to spend longer uninterrupted intervals exploring the strongly chaotic region before returning to sticky neighborhoods of regular islands. The small values of $1-R^2$ shown in Fig.~\ref{fig:exponents}(d) further indicate that the exponential description provides a robust characterization of the high--RTE episode statistics across the considered energy range. This is also consistent with the identification of $\kappa$ with an escape rate associated with the hyperbolic component of the chaotic saddle~\cite{Altmann2008, Altmann2009}, which decreases as that component occupies a larger portion of the accessible phase space. We have also verified that these results do not depend on the particular values chosen to define the episodes: repeating the whole analysis with thresholds $1.8$ and $3.2$, which also lie within the minimum separating the two modes of the finite--time RTE distribution, leaves both exponents in the same range and preserves both the strong oscillations of $\gamma$ with the energy and the overall decrease of $\kappa$. The individual values change, as expected when the criterion that defines an episode is modified, but the conclusions drawn from Fig.~\ref{fig:exponents} are unaffected.

    Finally, we compare the ability of the RTE to distinguish chaotic from regular trajectories with a widely used variational chaos indicator, namely the smaller alignment index (SALI)~\cite{Skokos2001,Skokos2004}. The SALI quantifies chaos through the evolution of two initially independent deviation vectors along the trajectory. For chaotic motion, the vectors tend to align exponentially fast with the most unstable direction, causing the SALI to decay toward zero, whereas for regular motion it fluctuates around nonzero values. Due to its effectiveness in distinguishing regular and chaotic dynamics, the SALI, as well as its generalization, the generalized alignment index (GALI), has been extensively tested and widely applied in Hamiltonian systems and nonlinear dynamics~\cite{Skokos2007, Skokos2008, Manos2012, Skokos2016, Sales2026}. For this reason, it is frequently adopted as a benchmark for chaos detection~\cite{Bazzani2023}. Its computation, however, requires the integration of both the orbit and the corresponding variational equations.

    Figure~\ref{fig:rhoc} shows the proportion of chaotic trajectories $\rho_c$ as a function of the energy, computed from an ensemble of $10^4$ randomly selected initial conditions using both the RTE and the SALI. To classify trajectories, we adopt the following thresholds: trajectories with $\RTE > 2.5$ are considered chaotic, while SALI trajectories satisfying $\mathrm{SALI}(t_f) < 10^{-10}$ at $t_f = 10^4$ are classified as chaotic. The agreement between the two methods is remarkably strong across the entire energy interval. Both indicators show the same overall trend: the proportion of chaotic trajectories increases monotonically with the energy, reflecting the progressive enlargement of the chaotic component of phase space as the regular regions shrink.
    
    The close overlap between the two curves demonstrates that the RTE provides a classification of regular and chaotic dynamics consistent with a well-established variational method. We emphasize that the purpose of this comparison is not to claim that the RTE outperforms the SALI. Rather, SALI is used here as a benchmark to validate the recurrence--based characterization. The main advantage of the RTE is methodological: unlike SALI and other variational indicators, it depends only on the trajectory time series and does not require the equations of motion to be differentiated or the Jacobian matrix to be computed. Therefore, the results of Fig.~\ref{fig:rhoc} indicate that the RTE can be used as a reliable alternative for distinguishing regular and chaotic dynamics while remaining applicable in situations where only temporal data are available.

    \section{The paradigm Hamiltonian}
    \label{sec:paradigm}

    We now use a second Hamiltonian model to verify that the RTE-based characterization applied above is not specific to the Hénon--Heiles system. For this purpose, we consider the paradigm Hamiltonian introduced by Escande~\cite{Escande1985}, a periodically driven one-degree-of-freedom Hamiltonian that describes the motion of a particle interacting with two longitudinal waves. This model has been used as a minimal setting for studying universal aspects of Hamiltonian stochasticity, including KAM-torus breakup, stochastic layers, resonance overlap, chaotic transport, and the transition to large--scale stochastic motion~\cite{Escande1985}. Its relevance comes from the fact that, despite its simple form, it contains two competing resonant terms and displays the typical mixed phase space structures of typical Hamiltonian systems. Since the system is explicitly time dependent, it has one and a half degree of freedom. Its stroboscopic dynamics therefore provides a two--dimensional area--preserving representation in which regular islands, chaotic layers, sticky regions, and mixed phase space structure can be directly visualized. This makes the paradigm Hamiltonian a natural second test case for the RTE.

    The model is defined by
    \begin{equation}
        \label{eq:ph}
        H(p,q,t) = \frac{p^{2}}{2} - M \cos q - A \cos\bigl(k(q-t)\bigr),
    \end{equation}
    where $q$ and $p$ are canonically conjugate variables, $M$ controls the amplitude of the static potential, $A$ controls the amplitude of the traveling-wave perturbation, and $k$ is the wave number of the perturbation. For $A=0$, the system reduces to the integrable pendulum Hamiltonian. For $A \neq 0$, the time-periodic perturbation breaks part of the regular structure and generates a mixed phase space. Therefore, by varying $A$, one can move from nearly regular dynamics to increasingly chaotic dynamics while keeping the model simple enough for a direct comparison with the Hénon--Heiles results.

    To analyze the dynamics, we use the stroboscopic map obtained by sampling the trajectory at integer multiples of the forcing period
    \begin{equation}
        T = \frac{2\pi}{k}.
    \end{equation}
    The resulting map is area preserving and provides a natural two--dimensional representation of the dynamics in the $(q,p)$ plane. The RTE is then computed from the sequence of stroboscopic points, in the same spirit as the PSS-based recurrence analysis used for the Hénon--Heiles system.

    \begin{figure*}[t!]
        \centering
        \includegraphics[width=\linewidth]{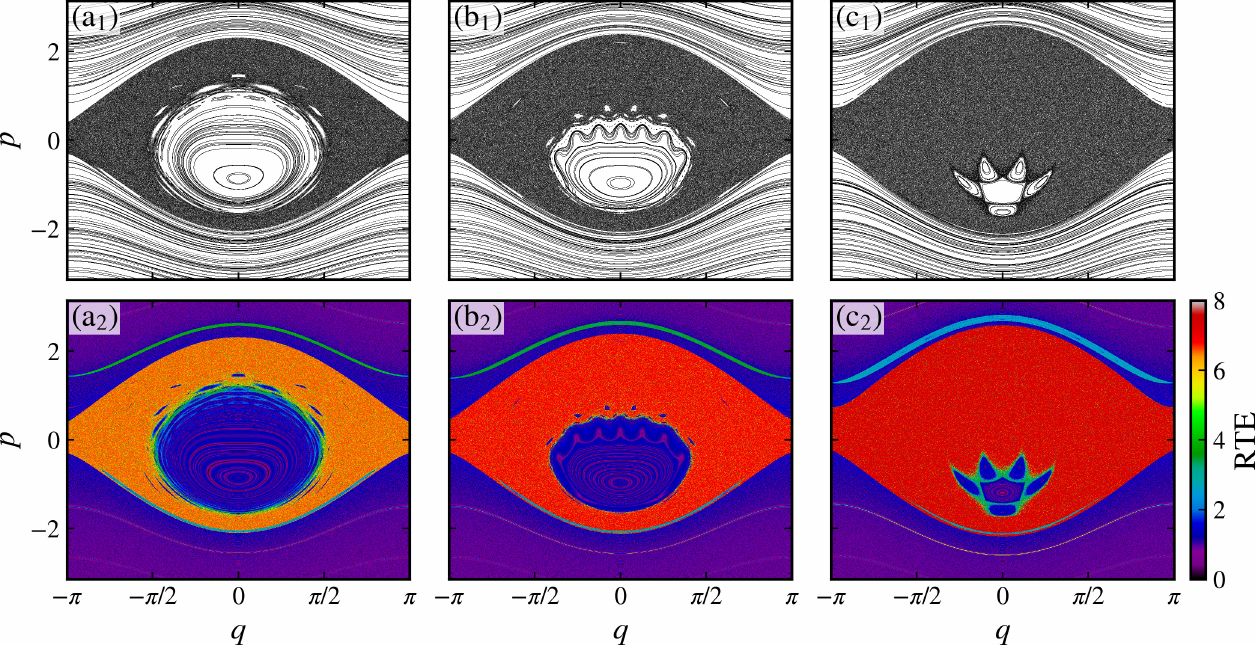}
        \caption{(Top row) The stroboscopic map sampled at multiples of the forcing period $T = 2\pi/k$ and (bottom row) the RTE calculated for $N_{\mathrm{cross}} = 10^4$ in a grid of initial conditions uniformly distributed in $(q, p)\in[-\pi, \pi]^2$ for the paradigm Hamiltonian [Eq.~\eqref{eq:ph}] with $M = 1.0$, $k = 1.0$, and (a) $A = 0.092$, (b) $A = 0.132$, and (c) $A = 0.248$.}
        \label{fig:PHRTE}
    \end{figure*}

    To integrate Eq.~\eqref{eq:ph} with a symplectic scheme over very long times, it is convenient to rewrite the problem as an autonomous Hamiltonian system in an extended phase space. Since the original Hamiltonian depends explicitly on time, it does not define an autonomous flow in the usual phase space $(q,p)$. As a consequence, the autonomous separable splitting used by the fourth--order Yoshida scheme is not directly available in the original $(q,p)$ phase space. The standard procedure is to promote time to a canonical coordinate.
    
    We introduce a new coordinate
    \begin{equation}
        \tau = t,
    \end{equation}
    and its canonically conjugate momentum $p_\tau$. The dynamics is then described in the extended phase space with canonical variables
    \begin{equation}
        \vb{Q} = (q,\tau), \qquad \vb{P} = (p, p_\tau),
    \end{equation}
    and with autonomous Hamiltonian
    \begin{equation}
        K(q,\tau,p,p_\tau) = \frac{p^{2}}{2} + p_\tau - M \cos q - A \cos\bigl(k(q-\tau)\bigr).
    \end{equation}
    The corresponding Hamilton equations are
    \begin{align}
        \dot{q} &= \frac{\partial K}{\partial p} = p, \\
        \dot{p} &= -\frac{\partial K}{\partial q} = -M \sin q - Ak \sin\bigl(k(q-\tau)\bigr), \\
        \dot{\tau} &= \frac{\partial K}{\partial p_\tau} = 1, \\
        \dot{p}_\tau &= -\frac{\partial K}{\partial \tau} = Ak \sin\bigl(k(q-\tau)\bigr).
    \end{align}
    It follows immediately that $\tau(t) = t + \tau_{0}$. Therefore, choosing $\tau_{0}=0$ recovers the original time dependence, and the nonautonomous system is exactly embedded into an autonomous Hamiltonian flow with two degrees of freedom.

    An important advantage of this formulation is that the extended Hamiltonian is separable. Indeed, it can be written as
    \begin{equation}
        K(\vb{Q}, \vb{P}) = T(\vb{P}) + V(\vb{Q}),
    \end{equation}
    where
    \begin{equation}
        T(\vb{P}) = \frac{p^{2}}{2} + p_\tau,
    \end{equation}
    and
    \begin{equation}
        V(\vb{Q}) = - M \cos q - A \cos\bigl(k(q-\tau)\bigr).
    \end{equation}
    Thus, the problem acquires the standard structure required by symplectic splitting integrators. This decomposition shows that the extended system can be integrated with the same symplectic algorithms used for autonomous separable Hamiltonians. In particular, second order Strang splitting, also known as the leapfrog scheme, can be directly employed, and higher order compositions such as the Yoshida integrator can be constructed from it. The essential point is that the explicit time dependence has been absorbed into the canonical pair $(\tau,p_\tau)$, restoring an autonomous symplectic structure in the extended phase space.
    

    Figure~\ref{fig:PHRTE} shows the stroboscopic map and the corresponding RTE computed on a grid of initial conditions in $(q,p)\in[-\pi,\pi]^2$ for three values of the perturbation amplitude $A$. The purpose of this example is not to repeat the full finite--time analysis performed for the Hénon--Heiles system, but to test whether the RTE also provides a meaningful recurrence--based description in a different Hamiltonian model.

    For the smallest perturbation, $A=0.092$ [Fig.~\ref{fig:PHRTE}(a)], the stroboscopic map in the top row is dominated by regular structures. A large central island is surrounded by a chaotic layer, while additional invariant curves persist near the outer region. The RTE field in the bottom row reproduces this organization clearly: inside the regular island and along invariant curves, the RTE assumes small values, whereas the chaotic layer is associated with larger values. Thus, as in the Hénon--Heiles system, low recurrence--time entropy identifies regular motion, while high recurrence--time entropy identifies chaotic transport.

    Increasing the perturbation to $A=0.132$ [Fig.~\ref{fig:PHRTE}(b)] enlarges the chaotic component and deforms the central regular region. The RTE map follows this change: the high--RTE region expands, while low--RTE values remain localized inside the surviving regular islands. Intermediate RTE values appear near the borders of the islands, where sticky dynamics is expected. This shows that the RTE is not merely separating regular and chaotic regions; it also reflects the transition layers generated by the mixed phase space structure.

    For the largest perturbation, $A=0.248$ [Fig.~\ref{fig:PHRTE}(c)], the chaotic sea occupies most of the central region, and only smaller regular islands remain visible. The corresponding RTE map is dominated by large values over the chaotic component, while the remaining islands and their immediate neighborhoods are marked by lower values. The agreement between the stroboscopic maps and the RTE fields confirms that the recurrence--time statistics recover the main geometrical structures of the phase space.

    Therefore, Fig.~\ref{fig:PHRTE} demonstrates that the RTE characterization is not specific to the Hénon--Heiles Hamiltonian. The same recurrence--based indicator identifies regular islands, chaotic seas, and sticky transition regions in a periodically driven Hamiltonian system. This supports the use of the RTE as a recurrence--based diagnostic for mixed Hamiltonian dynamics, both in autonomous systems with two degrees of freedom sampled on a PSS and in time-periodic systems with one degree of freedom represented by a stroboscopic map.

    \section{A three--degrees--of--freedom Hamiltonian system}
    \label{sec:threedof}

    The analysis presented so far has been restricted to systems whose surface of section is two--dimensional. In this section, we apply the same RTE--based characterization to a Hamiltonian system with three degrees of freedom, in order to verify whether the correspondence between the RTE and the largest Lyapunov exponent persists when the section becomes four--dimensional. We consider the following Hamiltonian~\cite{Contopoulos1985, Skokos2004}:
    \begin{equation}
        \label{eq:3dof}
        \begin{aligned}
            &H(x, y, z, p_x, p_y, p_z) = \frac{1}{2}\qty(p_x^2 + p_y^2 + p_z^2) +\\
            &\frac{1}{2}\qty(Ax^2 + By^2 + Cz^2) - \eta_1 xz^2 - \eta_2 yz^2,
        \end{aligned}
    \end{equation}
    which describes the motion in a galactic--type potential and has been used as a benchmark for chaos indicators in higher--dimensional Hamiltonian systems, in particular for the SALI~\cite{Skokos2004}. In all our numerical simulations, we keep the parameters fixed at $A = 0.9$, $B = 0.4$, $C = 0.225$, $\eta_1 = 0.56$, $\eta_2 = 0.2$, with total energy $E = 0.00765$, which are the same values considered in Ref.~\cite{Skokos2004}. The equations of motion are integrated with the same fourth--order Yoshida symplectic scheme and time step used in the previous sections.

    \begin{figure*}[t!]
        \centering
        \includegraphics[width=\linewidth]{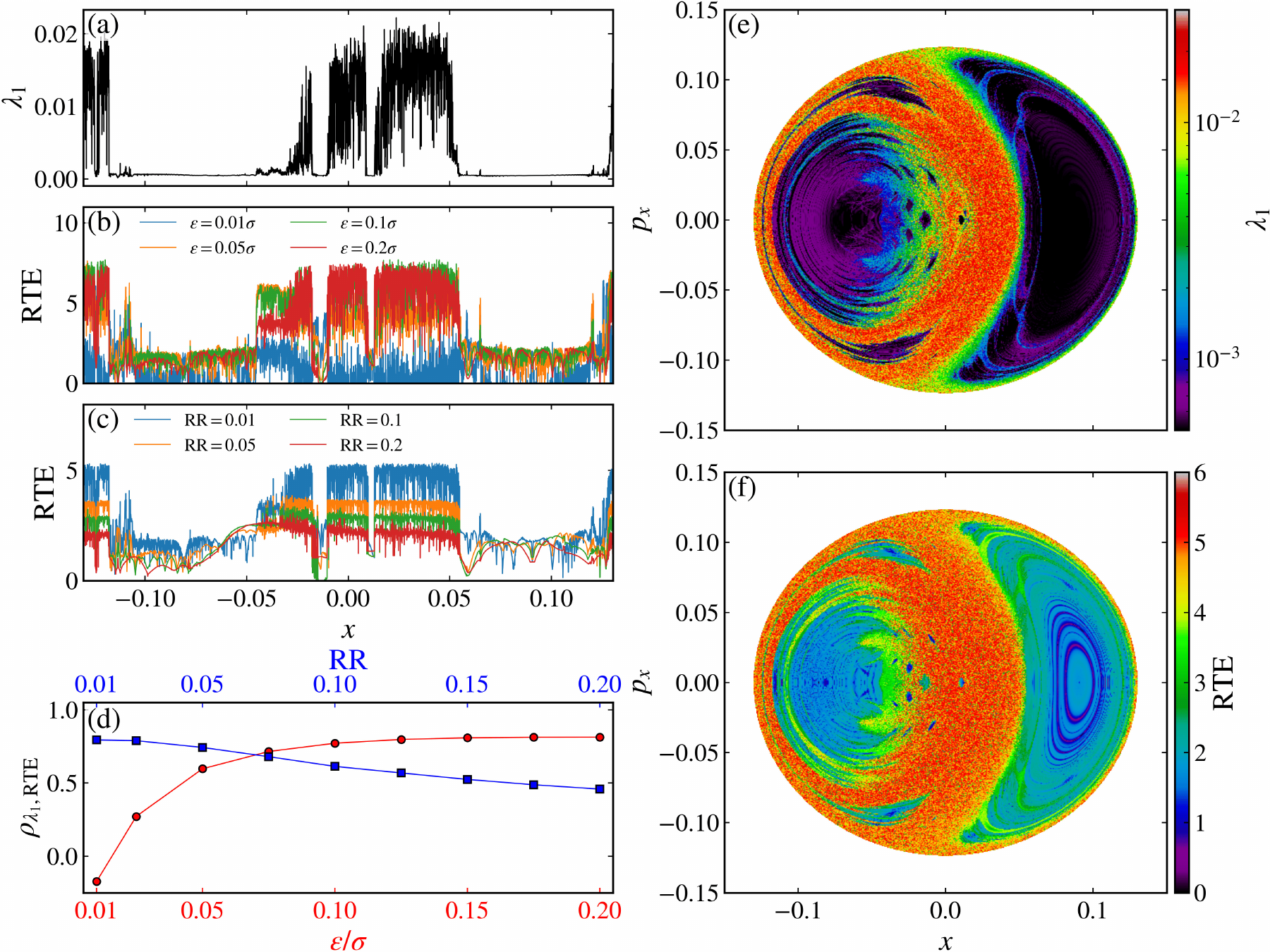}
        \caption{(a) The largest Lyapunov exponent, $\lambda_1$, for $t_{\mathrm{total}} = 10^4$, (b) and (c) the RTE for different threshold values based on the time series' standard deviation, $\sigma$, and based on a fixed recurrence rate, RR, respectively, for $N_{\mathrm{cross}} = 10^4$, as a function of $x$ for $y = 0$, $z = 0$, $p_x = 0$, $p_y = 0$, $p_z = p_z(x, y, z, p_x, p_y, E) > 0$, and $E = 0.00765$. (d) The Pearson correlation coefficient between $\lambda_1$ and the RTE for the two methods of determining $\varepsilon$. Each method has its own abscissa: the circles (red) are read on the bottom axis, which gives $\varepsilon$ in units of $\sigma$, and the squares (blue) are read on the top axis, which gives the imposed recurrence rate. The two quantities are distinct parameters that happen to span the same numerical range, and the abscissa must not be read as a common scale. (e) and (f) $\lambda_1$ for $t_{\mathrm{total}} = 10^4$ and RTE for $N_{\mathrm{cross}} = 10^4$ with $\mathrm{RR} = 0.01$, respectively, in a grid of initial conditions uniformly distributed in $(x, p_x) \in [-0.15, 0.15]^2$ with $y = 0$, $z = 0$, $p_y = 0$, $p_z = p_z(x, y, z, p_x, p_y, E) > 0$, and $E = 0.00765$.}
        \label{fig:3dof}
    \end{figure*}

    We compute the largest Lyapunov exponent, $\lambda_1$, [Fig.~\ref{fig:3dof}(a)] and the RTE [Figs.~\ref{fig:3dof}(b) and \ref{fig:3dof}(c)] as a function of $x$ for fixed $y = 0$, $z = 0$, $p_x = 0$, $p_y = 0$, and $p_z$ determined by the total energy, $p_z = p_z(x, y, z, p_x, p_y, E) > 0$. The recurrence matrix, in this case, is constructed by restricting the dynamics to the PSS defined by $z = 0$ and $p_z > 0$, and then considering the corresponding sequence of intersection points $(x_n, y_n, p_{x, n}, p_{y, n})$, i.e., the recurrence analysis is performed on the four--dimensional sequence generated by the successive crossings with the chosen surface of section. For such a multi--dimensional sequence, the threshold based on the standard deviation is defined using the largest standard deviation among the components of the sequence, and the distance between states is computed with the maximum norm, as in the previous sections.
    Similarly to what we have seen in Fig.~\ref{fig:LLERTE_vs_y}, both quantities show a clear correlation: large values of $\lambda_1$ are mirrored by large values of RTE, whereas the RTE is small wherever $\lambda_1 \approx 0$.

    The RTE values themselves depend on the threshold, but, as in the two--degrees--of--freedom case, the characterization of the dynamics does not. Figure~\ref{fig:3dof}(d) shows the Pearson correlation coefficient between $\lambda_1$ and the RTE for both methods, each read on its own abscissa, as in Fig.~\ref{fig:LLERTE_vs_y}(d): the circles are read on the bottom axis, which gives $\varepsilon/\sigma$, and the squares on the top axis, which gives RR. The correlation lies between approximately $0.6$ and $0.8$ for $\varepsilon$ between $0.05\sigma$ and $0.2\sigma$ and for RR between $0.01$ and $0.1$, that is, over a broad range of thresholds and for two entirely different ways of prescribing them. This is consistent with what has been reported for area--preserving maps, where the choice of $\varepsilon$ was also found not to be critical~\cite{Sales2023}. The only exception occurs at the smallest threshold, $\varepsilon = 0.01\sigma$, for which the correlation becomes slightly negative. This is a consequence of the dimension of the section. The recurrence rate obtained for a given threshold corresponds to the correlation sum of the sequence of crossings~\cite{Marwan2007}, and therefore, for a fixed threshold, the number of recurrent states decreases as the correlation dimension of the set explored by the trajectory in the section increases~\cite{Grassberger1983, Marwan2007}. Regular orbits lie on three--dimensional invariant tori, which intersect the section on a two--dimensional surface, whereas chaotic orbits fill a four--dimensional region. For the same $\varepsilon$, chaotic orbits therefore produce far fewer recurrent states than regular ones. If $\varepsilon$ is chosen too small, there are almost no recurrence points and nothing can be learned about the recurrence structure of the underlying system~\cite{Marwan2007}. In a four--dimensional section this happens first for the chaotic orbits, whose recurrence structure is essentially destroyed while the regular ones still recur, so that the RTE in the chaotic regions becomes comparable to the one obtained in the regular regions, as can be seen in Fig.~\ref{fig:3dof}(b). In a two--dimensional section this does not happen, since the difference in dimension between the sets explored by regular and chaotic orbits is smaller. The practical consequence is simply that the threshold should not be taken exceedingly small when the section is higher--dimensional, in accordance with the general observation that the choice of $\varepsilon$ depends strongly on the system under study~\cite{Marwan2007}. Fixing the recurrence rate ensures this automatically, since $\varepsilon$ is then adjusted for each trajectory so as to produce a prescribed number of recurrent states, which preserves the recurrence point density and allows the comparison of RPs of different systems without normalizing the time series beforehand~\cite{Marwan2007}, and it is for this reason that we have used $\mathrm{RR} = 0.01$ in Fig.~\ref{fig:3dof}(f), as an illustration that the characterization does not depend on the particular way in which the threshold is chosen.

    To better visualize the correlation between $\lambda_1$ and the RTE, we compute both quantities on a grid of uniformly distributed initial conditions in $(x, p_x) \in [-0.15, 0.15]^2$ with $y = 0$, $z = 0$, $p_y = 0$, and $p_z = p_z(x, y, z, p_x, p_y, E) > 0$. The results are shown in Figs.~\ref{fig:3dof}(e) and \ref{fig:3dof}(f). The regions with fully developed chaos, i.e., with large $\lambda_1$, also exhibit large RTE. Inside the regular regions, where $\lambda_1 \approx 0$, the RTE is small. Additionally, intermediate values of $\lambda_1$ correspond to intermediate values of the RTE. These results therefore demonstrate that the RTE characterization is not specific to systems with two degrees of freedom: it also identifies regular and chaotic regions in higher--dimensional systems.

    Finally, we note that the numerical values of the RTE depend on the dimension of the section and on the number of crossings considered, so the threshold $\RTE = 2.5$ adopted for the Hénon--Heiles system in Sec.~\ref{sec:hh_results} is not transferable to other systems and has to be recalibrated whenever the RTE is used as a binary chaos detector.

    \section{Conclusions}
    \label{sec:conclusions}

    In this paper, we have studied the use of the recurrence time entropy to characterize weakly chaotic dynamics in Hamiltonian systems with mixed phase space. The main goal has been to verify whether the RTE-based approach, previously used in area--preserving maps~\cite{Sales2023, Souza2024, Viana2025}, can also be applied to Hamiltonian flows when the dynamics is sampled on a Poincaré surface of section or through a stroboscopic map. For this purpose, we have considered the Hénon--Heiles Hamiltonian as a paradigmatic autonomous system with two degrees of freedom and the paradigm Hamiltonian as a periodically driven system with one degree of freedom.

    For the Hénon--Heiles system, we have shown that the RTE calculated from the sequence of PSS crossings reproduces the main phase space structures observed through the largest Lyapunov exponent. Regular islands have been associated with low RTE values, whereas chaotic regions have been associated with larger values. Near the boundaries of regular islands, the RTE has assumed intermediate values, indicating its sensitivity to sticky transition regions and weakly chaotic dynamics. The comparison with the LLE has also shown a strong positive correlation between the two quantities, confirming that the recurrence--time statistics have captured the same global separation between regular and chaotic motion.
    
    We have also analyzed the finite--time RTE along individual chaotic trajectories. The resulting time series have displayed strong fluctuations, with abrupt transitions between high--RTE and low--RTE episodes. These fluctuations have reflected the intermittent character of sticky dynamics: high--RTE intervals have corresponded to extended exploration of the chaotic component, whereas low--RTE intervals have been generated by temporary trapping near regular structures. By projecting the finite--time RTE values back onto the PSS, we have shown that the low-- and intermediate--RTE episodes have not been randomly distributed, but have been localized near regular islands and their surrounding sticky layers.
    
    The trapping--time analysis has further confirmed this interpretation. low--RTE episodes, associated with trapping near regular structures, have produced cumulative distributions with algebraic tails. In contrast, high--RTE episodes, associated with extended exploration of the chaotic component, have produced exponentially decaying distributions. Therefore, the finite--time RTE has not only identified where sticky dynamics occurs in phase space, but has also recovered the expected long-tailed statistics of trapping events in mixed Hamiltonian systems. By extending this analysis over a broad range of energies, we have shown that the trapping exponents vary nontrivially with the phase space structure. The power-law exponent associated with low--RTE episodes has exhibited strong oscillations as the energy changes, reflecting the sensitive dependence of sticky transport on the evolving hierarchical organization of the mixed phase space. By contrast, the exponential decay rate of high--RTE episodes has shown an overall decrease with energy, indicating that trajectories remain for progressively longer times in the strongly chaotic component as the chaotic sea expands.
    
    We have further compared the ability of the RTE to distinguish regular and chaotic trajectories with the Smaller Alignment Index (SALI), a well-established variational chaos indicator. The comparison has shown strong agreement between the two methods in the classification of chaotic motion across the explored energy range. In particular, both indicators have captured the monotonic increase in the proportion of chaotic trajectories with energy, consistent with the progressive enlargement of the chaotic component of phase space. This agreement has provided an additional validation of the RTE as an indicator of regular and chaotic motion.
    
    We have also applied the same recurrence--based analysis to the paradigm Hamiltonian. The RTE maps obtained from the stroboscopic dynamics have reproduced the main structures of the phase space for different perturbation amplitudes. Low RTE values have identified regular islands and invariant curves, while large RTE values have marked the chaotic component. This second example has shown that the method has not been specific to the Hénon--Heiles Hamiltonian, but can also be used in time-periodic Hamiltonian systems represented by stroboscopic maps.
    
    Finally, we have extended the analysis to a Hamiltonian system with three degrees of freedom, whose surface of section is four--dimensional. The RTE computed from the sequence of crossings has reproduced the separation between regular and chaotic regions obtained from the largest Lyapunov exponent, both along a one--dimensional scan of initial conditions and on a two--dimensional grid, which shows that the characterization is not restricted to systems with two degrees of freedom. For both this system and the Hénon--Heiles Hamiltonian we have also compared two different methods of determining the recurrence threshold, one based on the standard deviation of the time series and one based on a fixed recurrence rate, and the correlation with the largest Lyapunov exponent has remained high for both of them over a broad range of parameters. Taken together, the three systems considered here indicate that our conclusions do not rely on specific properties of the Hénon--Heiles Hamiltonian. What the analysis requires is a mixed phase space, in which regular islands coexist with a chaotic component and are surrounded by hierarchical structures and partial transport barriers, which is the generic situation in nonintegrable Hamiltonian systems.
    
    Overall, our results have shown that the RTE provides a useful recurrence--based diagnostic for mixed Hamiltonian dynamics that depends only on temporal series. It has distinguished regular and chaotic regions, detected sticky layers, resolved intermittent changes along chaotic trajectories, and recovered the different statistical behavior of trapping and chaotic-exploration episodes. Moreover, the comparison with the SALI has shown that the RTE reproduces the same global trend in the proportion of chaotic trajectories as the energy increases, supporting its use as a reliable indicator for distinguishing regular and chaotic motion. Unlike variational chaos indicators, however, the RTE does not require the integration of variational equations or the computation of the Jacobian matrix. These results support the use of the RTE as a complementary tool to Lyapunov and alignment--based indicators in the characterization of weak chaos in Hamiltonian systems. A systematic comparison between the RTE and other time series chaos detectors that are not derived from recurrence plots, such as the permutation entropy~\cite{Bandt2002}, is an interesting direction that we intend to pursue in a forthcoming work.

    \section*{Code availability}

    The source code to reproduce the results reported in this paper is available online in the GitHub repository: \href{https://github.com/mrolims-publications/weak-chaos-hamiltonian-flows}{https://github.com/mrolims-publications/weak-chaos-hamiltonian-flows}.
    
    \section*{Acknowledgments}
    
    This work was supported by the São Paulo Research Foundation (FAPESP, Brazil), under Grant Nos. 2023/08698-9, 2023/16146-6, 2024/05700-5, 2024/20417-8, and 2025/14544-0, by the National Council for Scientific and Technological Development (CNPq, Brazil), under Grant Nos.~304616/2021-4, 309670/2023-3, and 304398/2023-3, and by the Paraná Research Foundation (Fundação Araucária). The computational simulations in this research were supported by resources supplied by the Center for Scientific Computing (NCC/GridUNESP) of the São Paulo State University (UNESP) and by the \href{http://portal.if.usp.br/controle/}{Oscillations Control Group} of the University of São Paulo. We would like to thank the \href{https://www.105groupscience.com}{105 Group Science} for fruitful discussions. We would also like to acknowledge the use of the \textit{pynamicalsys} package~\cite{pynamicalsys}.

    \appendix

    \section{Dependence of the RTE on the length of the time series}
    \label{sec:appendix}

\begin{figure}[t!]
        \centering
        \includegraphics[width=\linewidth]{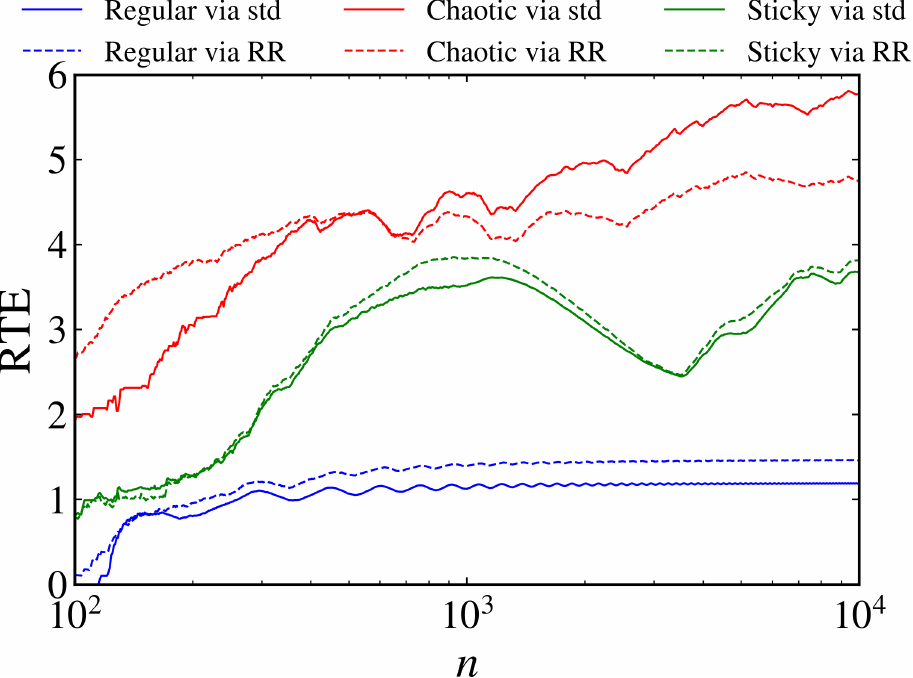}
        \caption{The RTE computed from the first $n$ crossings of the PSS for the three trajectories of Fig.~\ref{fig:recmats} at $E = 1/9$: regular (blue), sticky (green), and chaotic (red). The solid lines correspond to the threshold determined as a fraction of the time series' standard deviation, $\varepsilon = 0.05\sigma$, and the dashed lines to the threshold determined by fixing the recurrence rate at $\mathrm{RR} = 0.01$.}
        \label{fig:rte_vs_n}
    \end{figure}

    Throughout this paper the RTE has been computed from a fixed number of crossings of the PSS, $N_{\mathrm{cross}} = 10^4$. In this appendix we examine how the RTE depends on the length of the sequence used to construct the recurrence matrix. For this purpose, we consider the three trajectories of Fig.~\ref{fig:recmats}, namely the regular, the sticky and the chaotic ones, all at $E = 1/9$, and compute the RTE from the first $n$ crossings, with $n$ ranging from $10^2$ to $10^4$. The calculation is repeated for the two methods of determining the threshold discussed in Sec.~\ref{sec:rte}, $\varepsilon = 0.05\sigma$ and $\mathrm{RR} = 0.01$. The results are shown in Fig.~\ref{fig:rte_vs_n}.

    The regular and the chaotic trajectories behave in opposite ways. For the regular orbit, the RTE converges after a short transient and remains essentially constant from $n \approx 5\times 10^2$ onwards, at $1.20$ for the threshold based on the standard deviation and at $1.48$ for the one based on the recurrence rate. This is a consequence of the quasiperiodic nature of the motion: the number of distinct recurrence times of a quasiperiodic orbit is bounded, so that the distribution of white vertical lines has a bounded support and its entropy saturates. For the chaotic orbit, on the other hand, the RTE does not converge within the range considered, but increases slowly and irregularly, from about $2$ at $n = 10^2$ to $5.7$ at $n = 10^4$. In this case the recurrence times are broadly distributed and their support keeps growing as the orbit explores an increasingly large portion of the chaotic component, so that white vertical lines of increasing length are continuously added to the distribution.

    The sticky trajectory interpolates between these two behaviors, and its RTE is not even monotonic. For $n \lesssim 2\times 10^2$ it follows the regular curve, since the orbit is still trapped in the neighborhood of the regular structures. As the orbit escapes to the chaotic component, the RTE increases rapidly and approaches the chaotic curve, reaching a maximum near $n \approx 10^3$. It then decreases, down to approximately $2.5$ at $n \approx 3\times 10^3$, which corresponds to a second trapping episode: while the orbit is trapped, the recurrence times added to the distribution are similar to one another, the distribution becomes more concentrated and the entropy decreases. A subsequent escape makes the RTE increase again. The RTE computed from the first $n$ crossings therefore registers the alternation between trapping and exploration that is resolved in Sec.~\ref{sec:hh_results} by the finite--time RTE.

    Two conclusions follow. First, the numerical value of the RTE of a chaotic or sticky orbit depends on the length of the sequence, so that comparisons must be made at a fixed $n$. This is the reason why we keep $N_{\mathrm{cross}} = 10^4$ throughout the paper, and why a threshold value such as $\RTE = 2.5$ is tied to that choice. Second, and more importantly, the distinction between the different regimes is not affected by this dependence. For $n \gtrsim 2.5\times 10^2$ the three curves are well separated and ordered, with the regular orbit at the bottom, the sticky one in between and the chaotic one at the top, and this ordering is preserved over almost two decades in $n$. Since the regular curve saturates while the chaotic one keeps growing, the contrast between them in fact increases with $n$. The only limitation occurs for very short sequences, $n \lesssim 1.2\times 10^2$, for which the number of white vertical lines is too small for the distribution to be meaningful and the RTE of the regular orbit vanishes. Finally, the two methods lead to the same behavior in all three cases: they are almost indistinguishable for the sticky trajectory, differ by an approximately constant offset for the regular one, and separate appreciably only for the chaotic orbit at large $n$.


\begin{thebibliography}{78}%
\makeatletter
\providecommand \@ifxundefined [1]{%
 \@ifx{#1\undefined}
}%
\providecommand \@ifnum [1]{%
 \ifnum #1\expandafter \@firstoftwo
 \else \expandafter \@secondoftwo
 \fi
}%
\providecommand \@ifx [1]{%
 \ifx #1\expandafter \@firstoftwo
 \else \expandafter \@secondoftwo
 \fi
}%
\providecommand \natexlab [1]{#1}%
\providecommand \enquote  [1]{``#1''}%
\providecommand \bibnamefont  [1]{#1}%
\providecommand \bibfnamefont [1]{#1}%
\providecommand \citenamefont [1]{#1}%
\providecommand \href@noop [0]{\@secondoftwo}%
\providecommand \href [0]{\begingroup \@sanitize@url \@href}%
\providecommand \@href[1]{\@@startlink{#1}\@@href}%
\providecommand \@@href[1]{\endgroup#1\@@endlink}%
\providecommand \@sanitize@url [0]{\catcode `\\12\catcode `\$12\catcode `\&12\catcode `\#12\catcode `\^12\catcode `\_12\catcode `\%12\relax}%
\providecommand \@@startlink[1]{}%
\providecommand \@@endlink[0]{}%
\providecommand \url  [0]{\begingroup\@sanitize@url \@url }%
\providecommand \@url [1]{\endgroup\@href {#1}{\urlprefix }}%
\providecommand \urlprefix  [0]{URL }%
\providecommand \Eprint [0]{\href }%
\providecommand \doibase [0]{https://doi.org/}%
\providecommand \selectlanguage [0]{\@gobble}%
\providecommand \bibinfo  [0]{\@secondoftwo}%
\providecommand \bibfield  [0]{\@secondoftwo}%
\providecommand \translation [1]{[#1]}%
\providecommand \BibitemOpen [0]{}%
\providecommand \bibitemStop [0]{}%
\providecommand \bibitemNoStop [0]{.\EOS\space}%
\providecommand \EOS [0]{\spacefactor3000\relax}%
\providecommand \BibitemShut  [1]{\csname bibitem#1\endcsname}%
\let\auto@bib@innerbib\@empty
\bibitem [{\citenamefont {Lorenz}(1963)}]{Lorenz1963}%
  \BibitemOpen
  \bibfield  {author} {\bibinfo {author} {\bibfnamefont {E.~N.}\ \bibnamefont {Lorenz}},\ }\bibfield  {title} {\bibinfo {title} {Deterministic nonperiodic flow},\ }\href {https://doi.org/10.1175/1520-0469(1963)020<0130:DNF>2.0.CO;2} {\bibfield  {journal} {\bibinfo  {journal} {Journal of the Atmospheric Sciences}\ }\textbf {\bibinfo {volume} {20}},\ \bibinfo {pages} {130} (\bibinfo {year} {1963})}\BibitemShut {NoStop}%
\bibitem [{\citenamefont {R{\"o}ssler}(1976)}]{Rossler1976}%
  \BibitemOpen
  \bibfield  {author} {\bibinfo {author} {\bibfnamefont {O.~E.}\ \bibnamefont {R{\"o}ssler}},\ }\bibfield  {title} {\bibinfo {title} {An equation for continuous chaos},\ }\href {https://doi.org/10.1016/0375-9601(76)90101-8} {\bibfield  {journal} {\bibinfo  {journal} {Physics Letters A}\ }\textbf {\bibinfo {volume} {57}},\ \bibinfo {pages} {397} (\bibinfo {year} {1976})}\BibitemShut {NoStop}%
\bibitem [{\citenamefont {Matsumoto}(1984)}]{Matsumoto1984}%
  \BibitemOpen
  \bibfield  {author} {\bibinfo {author} {\bibfnamefont {T.}~\bibnamefont {Matsumoto}},\ }\bibfield  {title} {\bibinfo {title} {A chaotic attractor from {C}hua's circuit},\ }\href {https://doi.org/10.1109/TCS.1984.1085459} {\bibfield  {journal} {\bibinfo  {journal} {IEEE Transactions on Circuits and Systems}\ }\textbf {\bibinfo {volume} {31}},\ \bibinfo {pages} {1055} (\bibinfo {year} {1984})}\BibitemShut {NoStop}%
\bibitem [{\citenamefont {Lin}\ \emph {et~al.}(2026{\natexlab{a}})\citenamefont {Lin}, \citenamefont {Deng}, \citenamefont {Zhang},\ and\ \citenamefont {Min}}]{Lin2026}%
  \BibitemOpen
  \bibfield  {author} {\bibinfo {author} {\bibfnamefont {H.}~\bibnamefont {Lin}}, \bibinfo {author} {\bibfnamefont {X.}~\bibnamefont {Deng}}, \bibinfo {author} {\bibfnamefont {Y.}~\bibnamefont {Zhang}},\ and\ \bibinfo {author} {\bibfnamefont {G.}~\bibnamefont {Min}},\ }\bibfield  {title} {\bibinfo {title} {A second-order memristor method to construct memristive neural networks with multi-butterfly and multi-scroll dynamics},\ }\href {https://doi.org/10.1109/TCSI.2026.3663432} {\bibfield  {journal} {\bibinfo  {journal} {IEEE Transactions on Circuits and Systems I: Regular Papers}\ }\textbf {\bibinfo {volume} {73}},\ \bibinfo {pages} {4840} (\bibinfo {year} {2026}{\natexlab{a}})}\BibitemShut {NoStop}%
\bibitem [{\citenamefont {Lin}\ \emph {et~al.}(2026{\natexlab{b}})\citenamefont {Lin}, \citenamefont {Li}, \citenamefont {Zhang}, \citenamefont {Deng}, \citenamefont {Chen},\ and\ \citenamefont {Min}}]{Lin2026b}%
  \BibitemOpen
  \bibfield  {author} {\bibinfo {author} {\bibfnamefont {H.}~\bibnamefont {Lin}}, \bibinfo {author} {\bibfnamefont {Z.}~\bibnamefont {Li}}, \bibinfo {author} {\bibfnamefont {Y.}~\bibnamefont {Zhang}}, \bibinfo {author} {\bibfnamefont {X.}~\bibnamefont {Deng}}, \bibinfo {author} {\bibfnamefont {X.}~\bibnamefont {Chen}},\ and\ \bibinfo {author} {\bibfnamefont {G.}~\bibnamefont {Min}},\ }\bibfield  {title} {\bibinfo {title} {Memristive {CNN} with multi-butterfly attractors: Mathematical modeling, dynamics analysis and application in secure communication},\ }\href {https://doi.org/10.1016/j.chaos.2026.117905} {\bibfield  {journal} {\bibinfo  {journal} {Chaos, Solitons \& Fractals}\ }\textbf {\bibinfo {volume} {206}},\ \bibinfo {pages} {117905} (\bibinfo {year} {2026}{\natexlab{b}})}\BibitemShut {NoStop}%
\bibitem [{\citenamefont {Chen}\ \emph {et~al.}(2025)\citenamefont {Chen}, \citenamefont {Pan},\ and\ \citenamefont {Li}}]{Chen2025}%
  \BibitemOpen
  \bibfield  {author} {\bibinfo {author} {\bibfnamefont {M.}~\bibnamefont {Chen}}, \bibinfo {author} {\bibfnamefont {Z.}~\bibnamefont {Pan}},\ and\ \bibinfo {author} {\bibfnamefont {S.}~\bibnamefont {Li}},\ }\bibfield  {title} {\bibinfo {title} {A memristive neural network with hidden multi-butterfly dynamics: dynamics analysis, circuit implementation, and secure encryption application},\ }\href {https://doi.org/10.1088/1402-4896/ae1fb5} {\bibfield  {journal} {\bibinfo  {journal} {Physica Scripta}\ }\textbf {\bibinfo {volume} {100}},\ \bibinfo {pages} {115238} (\bibinfo {year} {2025})}\BibitemShut {NoStop}%
\bibitem [{\citenamefont {Zhang}\ \emph {et~al.}(2025)\citenamefont {Zhang}, \citenamefont {Li}, \citenamefont {Moroz}, \citenamefont {Huang},\ and\ \citenamefont {Liu}}]{Zhang2025}%
  \BibitemOpen
  \bibfield  {author} {\bibinfo {author} {\bibfnamefont {X.}~\bibnamefont {Zhang}}, \bibinfo {author} {\bibfnamefont {C.}~\bibnamefont {Li}}, \bibinfo {author} {\bibfnamefont {I.}~\bibnamefont {Moroz}}, \bibinfo {author} {\bibfnamefont {K.}~\bibnamefont {Huang}},\ and\ \bibinfo {author} {\bibfnamefont {Z.}~\bibnamefont {Liu}},\ }\bibfield  {title} {\bibinfo {title} {Non-bifurcation regulation of chaos in a memristive {H}opfield neural network},\ }\href {https://doi.org/10.1007/s11071-025-10949-z} {\bibfield  {journal} {\bibinfo  {journal} {Nonlinear Dynamics}\ }\textbf {\bibinfo {volume} {113}},\ \bibinfo {pages} {15487} (\bibinfo {year} {2025})}\BibitemShut {NoStop}%
\bibitem [{\citenamefont {Altmann}\ and\ \citenamefont {T\'el}(2008)}]{Altmann2008}%
  \BibitemOpen
  \bibfield  {author} {\bibinfo {author} {\bibfnamefont {E.~G.}\ \bibnamefont {Altmann}}\ and\ \bibinfo {author} {\bibfnamefont {T.}~\bibnamefont {T\'el}},\ }\bibfield  {title} {\bibinfo {title} {{P}oincar\'e recurrences from the perspective of transient chaos},\ }\href {https://doi.org/10.1103/PhysRevLett.100.174101} {\bibfield  {journal} {\bibinfo  {journal} {Phys. Rev. Lett.}\ }\textbf {\bibinfo {volume} {100}},\ \bibinfo {pages} {174101} (\bibinfo {year} {2008})}\BibitemShut {NoStop}%
\bibitem [{\citenamefont {Altmann}\ and\ \citenamefont {T\'el}(2009)}]{Altmann2009}%
  \BibitemOpen
  \bibfield  {author} {\bibinfo {author} {\bibfnamefont {E.~G.}\ \bibnamefont {Altmann}}\ and\ \bibinfo {author} {\bibfnamefont {T.}~\bibnamefont {T\'el}},\ }\bibfield  {title} {\bibinfo {title} {{P}oincar\'e recurrences and transient chaos in systems with leaks},\ }\href {https://doi.org/10.1103/PhysRevE.79.016204} {\bibfield  {journal} {\bibinfo  {journal} {Phys. Rev. E}\ }\textbf {\bibinfo {volume} {79}},\ \bibinfo {pages} {016204} (\bibinfo {year} {2009})}\BibitemShut {NoStop}%
\bibitem [{\citenamefont {Lichtenberg}\ and\ \citenamefont {Lieberman}(1992)}]{lichtenberg2013regular}%
  \BibitemOpen
  \bibfield  {author} {\bibinfo {author} {\bibfnamefont {A.~J.}\ \bibnamefont {Lichtenberg}}\ and\ \bibinfo {author} {\bibfnamefont {M.~A.}\ \bibnamefont {Lieberman}},\ }\href@noop {} {\emph {\bibinfo {title} {Regular and chaotic dynamics}}},\ \bibinfo {series} {Applied Mathematical Sciences}, Vol.~\bibinfo {volume} {38}\ (\bibinfo  {publisher} {Springer-Verlag},\ \bibinfo {year} {1992})\BibitemShut {NoStop}%
\bibitem [{\citenamefont {MacKay}\ \emph {et~al.}(1984)\citenamefont {MacKay}, \citenamefont {Meiss},\ and\ \citenamefont {Percival}}]{Mackay1984}%
  \BibitemOpen
  \bibfield  {author} {\bibinfo {author} {\bibfnamefont {R.~S.}\ \bibnamefont {MacKay}}, \bibinfo {author} {\bibfnamefont {J.~D.}\ \bibnamefont {Meiss}},\ and\ \bibinfo {author} {\bibfnamefont {I.~C.}\ \bibnamefont {Percival}},\ }\bibfield  {title} {\bibinfo {title} {Stochasticity and transport in {H}amiltonian systems},\ }\href {https://doi.org/10.1103/PhysRevLett.52.697} {\bibfield  {journal} {\bibinfo  {journal} {Phys. Rev. Lett.}\ }\textbf {\bibinfo {volume} {52}},\ \bibinfo {pages} {697} (\bibinfo {year} {1984})}\BibitemShut {NoStop}%
\bibitem [{\citenamefont {Mackay}\ \emph {et~al.}(1984)\citenamefont {Mackay}, \citenamefont {Meiss},\ and\ \citenamefont {Percival}}]{Mackay1984b}%
  \BibitemOpen
  \bibfield  {author} {\bibinfo {author} {\bibfnamefont {R.~S.}\ \bibnamefont {Mackay}}, \bibinfo {author} {\bibfnamefont {J.~D.}\ \bibnamefont {Meiss}},\ and\ \bibinfo {author} {\bibfnamefont {I.~C.}\ \bibnamefont {Percival}},\ }\bibfield  {title} {\bibinfo {title} {Transport in {H}amiltonian systems},\ }\href {https://doi.org/https://doi.org/10.1016/0167-2789(84)90270-7} {\bibfield  {journal} {\bibinfo  {journal} {Physica D: Nonlinear Phenomena}\ }\textbf {\bibinfo {volume} {13}},\ \bibinfo {pages} {55} (\bibinfo {year} {1984})}\BibitemShut {NoStop}%
\bibitem [{\citenamefont {Meiss}\ and\ \citenamefont {Ott}(1986)}]{Meiss1986}%
  \BibitemOpen
  \bibfield  {author} {\bibinfo {author} {\bibfnamefont {J.~D.}\ \bibnamefont {Meiss}}\ and\ \bibinfo {author} {\bibfnamefont {E.}~\bibnamefont {Ott}},\ }\bibfield  {title} {\bibinfo {title} {Markov tree model of transport in area-preserving maps},\ }\href {https://doi.org/https://doi.org/10.1016/0167-2789(86)90041-2} {\bibfield  {journal} {\bibinfo  {journal} {Physica D: Nonlinear Phenomena}\ }\textbf {\bibinfo {volume} {20}},\ \bibinfo {pages} {387} (\bibinfo {year} {1986})}\BibitemShut {NoStop}%
\bibitem [{\citenamefont {Contopoulos}(1971)}]{Contopoulos1971}%
  \BibitemOpen
  \bibfield  {author} {\bibinfo {author} {\bibfnamefont {G.}~\bibnamefont {Contopoulos}},\ }\bibfield  {title} {\bibinfo {title} {Orbits in highly perturbed dynamical systems. iii. nonperiodic orbits},\ }\href {https://doi.org/10.1086/111098} {\bibfield  {journal} {\bibinfo  {journal} {Astronomical Journal}\ }\textbf {\bibinfo {volume} {76}},\ \bibinfo {pages} {147} (\bibinfo {year} {1971})}\BibitemShut {NoStop}%
\bibitem [{\citenamefont {Karney}(1983)}]{Karney1983}%
  \BibitemOpen
  \bibfield  {author} {\bibinfo {author} {\bibfnamefont {C.~F.~F.}\ \bibnamefont {Karney}},\ }\bibfield  {title} {\bibinfo {title} {Long-time correlations in the stochastic regime},\ }\href {https://doi.org/https://doi.org/10.1016/0167-2789(83)90232-4} {\bibfield  {journal} {\bibinfo  {journal} {Physica D: Nonlinear Phenomena}\ }\textbf {\bibinfo {volume} {8}},\ \bibinfo {pages} {360} (\bibinfo {year} {1983})}\BibitemShut {NoStop}%
\bibitem [{\citenamefont {Meiss}\ \emph {et~al.}(1983)\citenamefont {Meiss}, \citenamefont {Cary}, \citenamefont {Grebogi}, \citenamefont {Crawford}, \citenamefont {Kaufman},\ and\ \citenamefont {Abarbanel}}]{Meiss1983}%
  \BibitemOpen
  \bibfield  {author} {\bibinfo {author} {\bibfnamefont {J.~D.}\ \bibnamefont {Meiss}}, \bibinfo {author} {\bibfnamefont {J.~R.}\ \bibnamefont {Cary}}, \bibinfo {author} {\bibfnamefont {C.}~\bibnamefont {Grebogi}}, \bibinfo {author} {\bibfnamefont {J.~D.}\ \bibnamefont {Crawford}}, \bibinfo {author} {\bibfnamefont {A.~N.}\ \bibnamefont {Kaufman}},\ and\ \bibinfo {author} {\bibfnamefont {H.~D.~I.}\ \bibnamefont {Abarbanel}},\ }\bibfield  {title} {\bibinfo {title} {Correlations of periodic, area-preserving maps},\ }\href {https://doi.org/https://doi.org/10.1016/0167-2789(83)90019-2} {\bibfield  {journal} {\bibinfo  {journal} {Physica D: Nonlinear Phenomena}\ }\textbf {\bibinfo {volume} {6}},\ \bibinfo {pages} {375} (\bibinfo {year} {1983})}\BibitemShut {NoStop}%
\bibitem [{\citenamefont {Efthymiopoulos}\ \emph {et~al.}(1997)\citenamefont {Efthymiopoulos}, \citenamefont {Contopoulos}, \citenamefont {Voglis},\ and\ \citenamefont {Dvorak}}]{Efthymiopoulos1997}%
  \BibitemOpen
  \bibfield  {author} {\bibinfo {author} {\bibfnamefont {C.}~\bibnamefont {Efthymiopoulos}}, \bibinfo {author} {\bibfnamefont {G.}~\bibnamefont {Contopoulos}}, \bibinfo {author} {\bibfnamefont {N.}~\bibnamefont {Voglis}},\ and\ \bibinfo {author} {\bibfnamefont {R.}~\bibnamefont {Dvorak}},\ }\bibfield  {title} {\bibinfo {title} {Stickiness and cantori},\ }\href {https://doi.org/10.1088/0305-4470/30/23/016} {\bibfield  {journal} {\bibinfo  {journal} {Journal of Physics A: Mathematical and General}\ }\textbf {\bibinfo {volume} {30}},\ \bibinfo {pages} {8167} (\bibinfo {year} {1997})}\BibitemShut {NoStop}%
\bibitem [{\citenamefont {Cristadoro}\ and\ \citenamefont {Ketzmerick}(2008)}]{Cristadoro2008}%
  \BibitemOpen
  \bibfield  {author} {\bibinfo {author} {\bibfnamefont {G.}~\bibnamefont {Cristadoro}}\ and\ \bibinfo {author} {\bibfnamefont {R.}~\bibnamefont {Ketzmerick}},\ }\bibfield  {title} {\bibinfo {title} {Universality of algebraic decays in {H}amiltonian systems},\ }\href {https://doi.org/10.1103/PhysRevLett.100.184101} {\bibfield  {journal} {\bibinfo  {journal} {Phys. Rev. Lett.}\ }\textbf {\bibinfo {volume} {100}},\ \bibinfo {pages} {184101} (\bibinfo {year} {2008})}\BibitemShut {NoStop}%
\bibitem [{\citenamefont {Contopoulos}\ and\ \citenamefont {Harsoula}(2008)}]{CONTOPOULOS2008}%
  \BibitemOpen
  \bibfield  {author} {\bibinfo {author} {\bibfnamefont {G.}~\bibnamefont {Contopoulos}}\ and\ \bibinfo {author} {\bibfnamefont {M.}~\bibnamefont {Harsoula}},\ }\bibfield  {title} {\bibinfo {title} {Stickiness in chaos},\ }\href {https://doi.org/10.1142/S0218127408022172} {\bibfield  {journal} {\bibinfo  {journal} {International Journal of Bifurcation and Chaos}\ }\textbf {\bibinfo {volume} {18}},\ \bibinfo {pages} {2929} (\bibinfo {year} {2008})}\BibitemShut {NoStop}%
\bibitem [{\citenamefont {Contopoulos}\ and\ \citenamefont {Harsoula}(2010)}]{Contopoulos2010}%
  \BibitemOpen
  \bibfield  {author} {\bibinfo {author} {\bibfnamefont {G.}~\bibnamefont {Contopoulos}}\ and\ \bibinfo {author} {\bibfnamefont {M.}~\bibnamefont {Harsoula}},\ }\bibfield  {title} {\bibinfo {title} {Stickiness effects in conservative systems},\ }\href {https://doi.org/10.1142/S0218127410026915} {\bibfield  {journal} {\bibinfo  {journal} {International Journal of Bifurcation and Chaos}\ }\textbf {\bibinfo {volume} {20}},\ \bibinfo {pages} {2005} (\bibinfo {year} {2010})}\BibitemShut {NoStop}%
\bibitem [{\citenamefont {Solomon}\ \emph {et~al.}(1993)\citenamefont {Solomon}, \citenamefont {Weeks},\ and\ \citenamefont {Swinney}}]{solomon1993}%
  \BibitemOpen
  \bibfield  {author} {\bibinfo {author} {\bibfnamefont {T.~H.}\ \bibnamefont {Solomon}}, \bibinfo {author} {\bibfnamefont {E.~R.}\ \bibnamefont {Weeks}},\ and\ \bibinfo {author} {\bibfnamefont {H.~L.}\ \bibnamefont {Swinney}},\ }\bibfield  {title} {\bibinfo {title} {Observation of anomalous diffusion and {L}\'evy flights in a two-dimensional rotating flow},\ }\href {https://doi.org/10.1103/PhysRevLett.71.3975} {\bibfield  {journal} {\bibinfo  {journal} {Phys. Rev. Lett.}\ }\textbf {\bibinfo {volume} {71}},\ \bibinfo {pages} {3975} (\bibinfo {year} {1993})}\BibitemShut {NoStop}%
\bibitem [{\citenamefont {Contopoulos}(2002)}]{Contopoulos2002}%
  \BibitemOpen
  \bibfield  {author} {\bibinfo {author} {\bibfnamefont {G.}~\bibnamefont {Contopoulos}},\ }\bibinfo {title} {Order and chaos in general},\ in\ \href {https://doi.org/10.1007/978-3-662-04917-4_2} {\emph {\bibinfo {booktitle} {Order and Chaos in Dynamical Astronomy}}}\ (\bibinfo  {publisher} {Springer Berlin Heidelberg},\ \bibinfo {address} {Berlin, Heidelberg},\ \bibinfo {year} {2002})\ pp.\ \bibinfo {pages} {11--376}\BibitemShut {NoStop}%
\bibitem [{\citenamefont {Szezech}\ \emph {et~al.}(2005)\citenamefont {Szezech}, \citenamefont {Lopes},\ and\ \citenamefont {Viana}}]{Szezech2005}%
  \BibitemOpen
  \bibfield  {author} {\bibinfo {author} {\bibfnamefont {J.~D.}\ \bibnamefont {Szezech}}, \bibinfo {author} {\bibfnamefont {S.~R.}\ \bibnamefont {Lopes}},\ and\ \bibinfo {author} {\bibfnamefont {R.~L.}\ \bibnamefont {Viana}},\ }\bibfield  {title} {\bibinfo {title} {Finite-time {L}yapunov spectrum for chaotic orbits of non-integrable {H}amiltonian systems},\ }\href {https://doi.org/https://doi.org/10.1016/j.physleta.2004.12.058} {\bibfield  {journal} {\bibinfo  {journal} {Physics Letters A}\ }\textbf {\bibinfo {volume} {335}},\ \bibinfo {pages} {394} (\bibinfo {year} {2005})}\BibitemShut {NoStop}%
\bibitem [{\citenamefont {Harle}\ and\ \citenamefont {Feudel}(2007)}]{Harle2007}%
  \BibitemOpen
  \bibfield  {author} {\bibinfo {author} {\bibfnamefont {M.}~\bibnamefont {Harle}}\ and\ \bibinfo {author} {\bibfnamefont {U.}~\bibnamefont {Feudel}},\ }\bibfield  {title} {\bibinfo {title} {Hierarchy of islands in conservative systems yields multimodal distributions of {FTLE}s},\ }\href {https://doi.org/https://doi.org/10.1016/j.chaos.2005.09.031} {\bibfield  {journal} {\bibinfo  {journal} {Chaos, Solitons \& Fractals}\ }\textbf {\bibinfo {volume} {31}},\ \bibinfo {pages} {130} (\bibinfo {year} {2007})}\BibitemShut {NoStop}%
\bibitem [{\citenamefont {da~Silva}\ \emph {et~al.}(2015)\citenamefont {da~Silva}, \citenamefont {Manchein}, \citenamefont {Beims},\ and\ \citenamefont {Altmann}}]{Silva2015}%
  \BibitemOpen
  \bibfield  {author} {\bibinfo {author} {\bibfnamefont {R.~M.}\ \bibnamefont {da~Silva}}, \bibinfo {author} {\bibfnamefont {C.}~\bibnamefont {Manchein}}, \bibinfo {author} {\bibfnamefont {M.~W.}\ \bibnamefont {Beims}},\ and\ \bibinfo {author} {\bibfnamefont {E.~G.}\ \bibnamefont {Altmann}},\ }\bibfield  {title} {\bibinfo {title} {Characterizing weak chaos using time series of {L}yapunov exponents},\ }\href {https://doi.org/10.1103/PhysRevE.91.062907} {\bibfield  {journal} {\bibinfo  {journal} {Phys. Rev. E}\ }\textbf {\bibinfo {volume} {91}},\ \bibinfo {pages} {062907} (\bibinfo {year} {2015})}\BibitemShut {NoStop}%
\bibitem [{\citenamefont {Santos}\ \emph {et~al.}(2019)\citenamefont {Santos}, \citenamefont {Mugnaine}, \citenamefont {Szezech}, \citenamefont {Batista}, \citenamefont {Caldas},\ and\ \citenamefont {Viana}}]{Santos2019}%
  \BibitemOpen
  \bibfield  {author} {\bibinfo {author} {\bibfnamefont {M.~S.}\ \bibnamefont {Santos}}, \bibinfo {author} {\bibfnamefont {M.}~\bibnamefont {Mugnaine}}, \bibinfo {author} {\bibfnamefont {J.}~\bibnamefont {Szezech}, \bibfnamefont {José~D.}}, \bibinfo {author} {\bibfnamefont {A.~M.}\ \bibnamefont {Batista}}, \bibinfo {author} {\bibfnamefont {I.~L.}\ \bibnamefont {Caldas}},\ and\ \bibinfo {author} {\bibfnamefont {R.~L.}\ \bibnamefont {Viana}},\ }\bibfield  {title} {\bibinfo {title} {Using rotation number to detect sticky orbits in {H}amiltonian systems},\ }\href {https://doi.org/10.1063/1.5078533} {\bibfield  {journal} {\bibinfo  {journal} {Chaos: An Interdisciplinary Journal of Nonlinear Science}\ }\textbf {\bibinfo {volume} {29}},\ \bibinfo {pages} {043125} (\bibinfo {year} {2019})}\BibitemShut {NoStop}%
\bibitem [{\citenamefont {Borin}(2024)}]{Borin2024}%
  \BibitemOpen
  \bibfield  {author} {\bibinfo {author} {\bibfnamefont {D.}~\bibnamefont {Borin}},\ }\bibfield  {title} {\bibinfo {title} {Hurst exponent: A method for characterizing dynamical traps},\ }\href {https://doi.org/10.1103/PhysRevE.110.064227} {\bibfield  {journal} {\bibinfo  {journal} {Phys. Rev. E}\ }\textbf {\bibinfo {volume} {110}},\ \bibinfo {pages} {064227} (\bibinfo {year} {2024})}\BibitemShut {NoStop}%
\bibitem [{\citenamefont {Borin}\ \emph {et~al.}(2025)\citenamefont {Borin}, \citenamefont {Szezech},\ and\ \citenamefont {Sales}}]{Borin2025}%
  \BibitemOpen
  \bibfield  {author} {\bibinfo {author} {\bibfnamefont {D.}~\bibnamefont {Borin}}, \bibinfo {author} {\bibfnamefont {J.~D.}\ \bibnamefont {Szezech}},\ and\ \bibinfo {author} {\bibfnamefont {M.~R.}\ \bibnamefont {Sales}},\ }\bibfield  {title} {\bibinfo {title} {Characterizing and quantifying weak chaos in fractional dynamics},\ }\href {https://doi.org/https://doi.org/10.1016/j.chaos.2025.117137} {\bibfield  {journal} {\bibinfo  {journal} {Chaos, Solitons \& Fractals}\ }\textbf {\bibinfo {volume} {200}},\ \bibinfo {pages} {117137} (\bibinfo {year} {2025})}\BibitemShut {NoStop}%
\bibitem [{\citenamefont {Altmann}\ \emph {et~al.}(2005)\citenamefont {Altmann}, \citenamefont {Motter},\ and\ \citenamefont {Kantz}}]{Altmann2005}%
  \BibitemOpen
  \bibfield  {author} {\bibinfo {author} {\bibfnamefont {E.~G.}\ \bibnamefont {Altmann}}, \bibinfo {author} {\bibfnamefont {A.~E.}\ \bibnamefont {Motter}},\ and\ \bibinfo {author} {\bibfnamefont {H.}~\bibnamefont {Kantz}},\ }\bibfield  {title} {\bibinfo {title} {Stickiness in mushroom billiards},\ }\href {https://doi.org/10.1063/1.1979211} {\bibfield  {journal} {\bibinfo  {journal} {Chaos: An Interdisciplinary Journal of Nonlinear Science}\ }\textbf {\bibinfo {volume} {15}},\ \bibinfo {pages} {033105} (\bibinfo {year} {2005})}\BibitemShut {NoStop}%
\bibitem [{\citenamefont {Altmann}\ \emph {et~al.}(2006)\citenamefont {Altmann}, \citenamefont {Motter},\ and\ \citenamefont {Kantz}}]{Altmann2006}%
  \BibitemOpen
  \bibfield  {author} {\bibinfo {author} {\bibfnamefont {E.~G.}\ \bibnamefont {Altmann}}, \bibinfo {author} {\bibfnamefont {A.~E.}\ \bibnamefont {Motter}},\ and\ \bibinfo {author} {\bibfnamefont {H.}~\bibnamefont {Kantz}},\ }\bibfield  {title} {\bibinfo {title} {Stickiness in {H}amiltonian systems: From sharply divided to hierarchical phase space},\ }\href {https://doi.org/10.1103/PhysRevE.73.026207} {\bibfield  {journal} {\bibinfo  {journal} {Phys. Rev. E}\ }\textbf {\bibinfo {volume} {73}},\ \bibinfo {pages} {026207} (\bibinfo {year} {2006})}\BibitemShut {NoStop}%
\bibitem [{\citenamefont {Abud}\ and\ \citenamefont {de~Carvalho}(2013)}]{Abud2013}%
  \BibitemOpen
  \bibfield  {author} {\bibinfo {author} {\bibfnamefont {C.~V.}\ \bibnamefont {Abud}}\ and\ \bibinfo {author} {\bibfnamefont {R.~E.}\ \bibnamefont {de~Carvalho}},\ }\bibfield  {title} {\bibinfo {title} {Multifractality, stickiness, and recurrence-time statistics},\ }\href {https://doi.org/10.1103/PhysRevE.88.042922} {\bibfield  {journal} {\bibinfo  {journal} {Phys. Rev. E}\ }\textbf {\bibinfo {volume} {88}},\ \bibinfo {pages} {042922} (\bibinfo {year} {2013})}\BibitemShut {NoStop}%
\bibitem [{\citenamefont {Lozej}(2020)}]{Lozej2020}%
  \BibitemOpen
  \bibfield  {author} {\bibinfo {author} {\bibfnamefont {C.}~\bibnamefont {Lozej}},\ }\bibfield  {title} {\bibinfo {title} {Stickiness in generic low-dimensional {H}amiltonian systems: A recurrence-time statistics approach},\ }\href {https://doi.org/10.1103/PhysRevE.101.052204} {\bibfield  {journal} {\bibinfo  {journal} {Phys. Rev. E}\ }\textbf {\bibinfo {volume} {101}},\ \bibinfo {pages} {052204} (\bibinfo {year} {2020})}\BibitemShut {NoStop}%
\bibitem [{\citenamefont {Zou}\ \emph {et~al.}(2007{\natexlab{a}})\citenamefont {Zou}, \citenamefont {Thiel}, \citenamefont {Romano},\ and\ \citenamefont {Kurths}}]{Zou2007b}%
  \BibitemOpen
  \bibfield  {author} {\bibinfo {author} {\bibfnamefont {Y.}~\bibnamefont {Zou}}, \bibinfo {author} {\bibfnamefont {M.}~\bibnamefont {Thiel}}, \bibinfo {author} {\bibfnamefont {M.~C.}\ \bibnamefont {Romano}},\ and\ \bibinfo {author} {\bibfnamefont {J.}~\bibnamefont {Kurths}},\ }\bibfield  {title} {\bibinfo {title} {{Characterization of stickiness by means of recurrence}},\ }\href {https://doi.org/10.1063/1.2785159} {\bibfield  {journal} {\bibinfo  {journal} {Chaos: An Interdisciplinary Journal of Nonlinear Science}\ }\textbf {\bibinfo {volume} {17}},\ \bibinfo {pages} {043101} (\bibinfo {year} {2007}{\natexlab{a}})}\BibitemShut {NoStop}%
\bibitem [{\citenamefont {Zou}\ \emph {et~al.}(2007{\natexlab{b}})\citenamefont {Zou}, \citenamefont {Paz\'o}, \citenamefont {Romano}, \citenamefont {Thiel},\ and\ \citenamefont {Kurths}}]{Zou2007}%
  \BibitemOpen
  \bibfield  {author} {\bibinfo {author} {\bibfnamefont {Y.}~\bibnamefont {Zou}}, \bibinfo {author} {\bibfnamefont {D.}~\bibnamefont {Paz\'o}}, \bibinfo {author} {\bibfnamefont {M.~C.}\ \bibnamefont {Romano}}, \bibinfo {author} {\bibfnamefont {M.}~\bibnamefont {Thiel}},\ and\ \bibinfo {author} {\bibfnamefont {J.}~\bibnamefont {Kurths}},\ }\bibfield  {title} {\bibinfo {title} {Distinguishing quasiperiodic dynamics from chaos in short-time series},\ }\href {https://doi.org/10.1103/PhysRevE.76.016210} {\bibfield  {journal} {\bibinfo  {journal} {Phys. Rev. E}\ }\textbf {\bibinfo {volume} {76}},\ \bibinfo {pages} {016210} (\bibinfo {year} {2007}{\natexlab{b}})}\BibitemShut {NoStop}%
\bibitem [{\citenamefont {Palmero}\ \emph {et~al.}(2022)\citenamefont {Palmero}, \citenamefont {Caldas},\ and\ \citenamefont {Sokolov}}]{Palmero2022}%
  \BibitemOpen
  \bibfield  {author} {\bibinfo {author} {\bibfnamefont {M.~S.}\ \bibnamefont {Palmero}}, \bibinfo {author} {\bibfnamefont {I.~L.}\ \bibnamefont {Caldas}},\ and\ \bibinfo {author} {\bibfnamefont {I.~M.}\ \bibnamefont {Sokolov}},\ }\bibfield  {title} {\bibinfo {title} {Finite-time recurrence analysis of chaotic trajectories in {H}amiltonian systems},\ }\href {https://doi.org/10.1063/5.0102424} {\bibfield  {journal} {\bibinfo  {journal} {Chaos}\ }\textbf {\bibinfo {volume} {32}},\ \bibinfo {pages} {113144} (\bibinfo {year} {2022})}\BibitemShut {NoStop}%
\bibitem [{\citenamefont {Eckmann}\ \emph {et~al.}(1987)\citenamefont {Eckmann}, \citenamefont {Kamphorst},\ and\ \citenamefont {Ruelle}}]{Eckmann1987}%
  \BibitemOpen
  \bibfield  {author} {\bibinfo {author} {\bibfnamefont {J.~P.}\ \bibnamefont {Eckmann}}, \bibinfo {author} {\bibfnamefont {S.~O.}\ \bibnamefont {Kamphorst}},\ and\ \bibinfo {author} {\bibfnamefont {D.}~\bibnamefont {Ruelle}},\ }\bibfield  {title} {\bibinfo {title} {Recurrence plots of dynamical systems},\ }\href {https://doi.org/10.1209/0295-5075/4/9/004} {\bibfield  {journal} {\bibinfo  {journal} {Europhysics Letters}\ }\textbf {\bibinfo {volume} {4}},\ \bibinfo {pages} {973} (\bibinfo {year} {1987})}\BibitemShut {NoStop}%
\bibitem [{\citenamefont {Marwan}\ \emph {et~al.}(2007)\citenamefont {Marwan}, \citenamefont {{Carmen Romano}}, \citenamefont {Thiel},\ and\ \citenamefont {Kurths}}]{Marwan2007}%
  \BibitemOpen
  \bibfield  {author} {\bibinfo {author} {\bibfnamefont {N.}~\bibnamefont {Marwan}}, \bibinfo {author} {\bibfnamefont {M.}~\bibnamefont {{Carmen Romano}}}, \bibinfo {author} {\bibfnamefont {M.}~\bibnamefont {Thiel}},\ and\ \bibinfo {author} {\bibfnamefont {J.}~\bibnamefont {Kurths}},\ }\bibfield  {title} {\bibinfo {title} {Recurrence plots for the analysis of complex systems},\ }\href {https://doi.org/https://doi.org/10.1016/j.physrep.2006.11.001} {\bibfield  {journal} {\bibinfo  {journal} {Physics Reports}\ }\textbf {\bibinfo {volume} {438}},\ \bibinfo {pages} {237} (\bibinfo {year} {2007})}\BibitemShut {NoStop}%
\bibitem [{\citenamefont {Sales}\ \emph {et~al.}(2023)\citenamefont {Sales}, \citenamefont {Mugnaine}, \citenamefont {{Szezech Jr.}}, \citenamefont {Viana}, \citenamefont {Caldas}, \citenamefont {Marwan},\ and\ \citenamefont {Kurths}}]{Sales2023}%
  \BibitemOpen
  \bibfield  {author} {\bibinfo {author} {\bibfnamefont {M.~R.}\ \bibnamefont {Sales}}, \bibinfo {author} {\bibfnamefont {M.}~\bibnamefont {Mugnaine}}, \bibinfo {author} {\bibfnamefont {J.~D.}\ \bibnamefont {{Szezech Jr.}}}, \bibinfo {author} {\bibfnamefont {R.~L.}\ \bibnamefont {Viana}}, \bibinfo {author} {\bibfnamefont {I.~L.}\ \bibnamefont {Caldas}}, \bibinfo {author} {\bibfnamefont {N.}~\bibnamefont {Marwan}},\ and\ \bibinfo {author} {\bibfnamefont {J.}~\bibnamefont {Kurths}},\ }\bibfield  {title} {\bibinfo {title} {{Stickiness and recurrence plots: An entropy-based approach}},\ }\href {https://doi.org/10.1063/5.0140613} {\bibfield  {journal} {\bibinfo  {journal} {Chaos: An Interdisciplinary Journal of Nonlinear Science}\ }\textbf {\bibinfo {volume} {33}},\ \bibinfo {pages} {033140} (\bibinfo {year} {2023})}\BibitemShut {NoStop}%
\bibitem [{\citenamefont {Daquin}\ and\ \citenamefont {Kov{\'a}cs}(2026)}]{Daquin2026}%
  \BibitemOpen
  \bibfield  {author} {\bibinfo {author} {\bibfnamefont {J.}~\bibnamefont {Daquin}}\ and\ \bibinfo {author} {\bibfnamefont {T.}~\bibnamefont {Kov{\'a}cs}},\ }\bibfield  {title} {\bibinfo {title} {Leveraging temporal features of the divergence quantifier of recurrence plot to detect chaos in conservative systems},\ }\bibfield  {journal} {\bibinfo  {journal} {The European Physical Journal Special Topics}\ }\href {https://doi.org/10.1140/epjs/s11734-026-02183-4} {10.1140/epjs/s11734-026-02183-4} (\bibinfo {year} {2026})\BibitemShut {NoStop}%
\bibitem [{\citenamefont {Little}\ \emph {et~al.}(2007)\citenamefont {Little}, \citenamefont {McSharry}, \citenamefont {Roberts}, \citenamefont {Costello},\ and\ \citenamefont {Moroz}}]{Little2007}%
  \BibitemOpen
  \bibfield  {author} {\bibinfo {author} {\bibfnamefont {M.~A.}\ \bibnamefont {Little}}, \bibinfo {author} {\bibfnamefont {P.~E.}\ \bibnamefont {McSharry}}, \bibinfo {author} {\bibfnamefont {S.~J.}\ \bibnamefont {Roberts}}, \bibinfo {author} {\bibfnamefont {D.~A.}\ \bibnamefont {Costello}},\ and\ \bibinfo {author} {\bibfnamefont {I.~M.}\ \bibnamefont {Moroz}},\ }\bibfield  {title} {\bibinfo {title} {Exploiting nonlinear recurrence and fractal scaling properties for voice disorder detection},\ }\href {https://doi.org/10.1186/1475-925X-6-23} {\bibfield  {journal} {\bibinfo  {journal} {BioMedical Engineering OnLine}\ }\textbf {\bibinfo {volume} {6}},\ \bibinfo {pages} {23} (\bibinfo {year} {2007})}\BibitemShut {NoStop}%
\bibitem [{\citenamefont {Baptista}\ \emph {et~al.}(2010)\citenamefont {Baptista}, \citenamefont {Ngamga}, \citenamefont {Pinto}, \citenamefont {Brito},\ and\ \citenamefont {Kurths}}]{Baptista2010}%
  \BibitemOpen
  \bibfield  {author} {\bibinfo {author} {\bibfnamefont {M.~S.}\ \bibnamefont {Baptista}}, \bibinfo {author} {\bibfnamefont {E.~J.}\ \bibnamefont {Ngamga}}, \bibinfo {author} {\bibfnamefont {P.~R.~F.}\ \bibnamefont {Pinto}}, \bibinfo {author} {\bibfnamefont {M.}~\bibnamefont {Brito}},\ and\ \bibinfo {author} {\bibfnamefont {J.}~\bibnamefont {Kurths}},\ }\bibfield  {title} {\bibinfo {title} {Kolmogorov–{S}inai entropy from recurrence times},\ }\href {https://doi.org/https://doi.org/10.1016/j.physleta.2009.12.057} {\bibfield  {journal} {\bibinfo  {journal} {Physics Letters A}\ }\textbf {\bibinfo {volume} {374}},\ \bibinfo {pages} {1135} (\bibinfo {year} {2010})}\BibitemShut {NoStop}%
\bibitem [{\citenamefont {Souza}\ \emph {et~al.}(2024)\citenamefont {Souza}, \citenamefont {Sales}, \citenamefont {Mugnaine}, \citenamefont {Szezech}, \citenamefont {Caldas},\ and\ \citenamefont {Viana}}]{Souza2024}%
  \BibitemOpen
  \bibfield  {author} {\bibinfo {author} {\bibfnamefont {L.~C.}\ \bibnamefont {Souza}}, \bibinfo {author} {\bibfnamefont {M.~R.}\ \bibnamefont {Sales}}, \bibinfo {author} {\bibfnamefont {M.}~\bibnamefont {Mugnaine}}, \bibinfo {author} {\bibfnamefont {J.~D.}\ \bibnamefont {Szezech}}, \bibinfo {author} {\bibfnamefont {I.~L.}\ \bibnamefont {Caldas}},\ and\ \bibinfo {author} {\bibfnamefont {R.~L.}\ \bibnamefont {Viana}},\ }\bibfield  {title} {\bibinfo {title} {Chaotic escape of impurities and sticky orbits in toroidal plasmas},\ }\href {https://doi.org/10.1103/PhysRevE.109.015202} {\bibfield  {journal} {\bibinfo  {journal} {Phys. Rev. E}\ }\textbf {\bibinfo {volume} {109}},\ \bibinfo {pages} {015202} (\bibinfo {year} {2024})}\BibitemShut {NoStop}%
\bibitem [{\citenamefont {Viana}\ \emph {et~al.}(2025)\citenamefont {Viana}, \citenamefont {Souza}, \citenamefont {Sales}, \citenamefont {Mugnaine}, \citenamefont {Szezech}, \citenamefont {Caldas}, \citenamefont {Marwan},\ and\ \citenamefont {Kurths}}]{Viana2025}%
  \BibitemOpen
  \bibfield  {author} {\bibinfo {author} {\bibfnamefont {R.~L.}\ \bibnamefont {Viana}}, \bibinfo {author} {\bibfnamefont {L.~C.}\ \bibnamefont {Souza}}, \bibinfo {author} {\bibfnamefont {M.~R.}\ \bibnamefont {Sales}}, \bibinfo {author} {\bibfnamefont {M.}~\bibnamefont {Mugnaine}}, \bibinfo {author} {\bibfnamefont {J.~D.}\ \bibnamefont {Szezech}}, \bibinfo {author} {\bibfnamefont {I.~L.}\ \bibnamefont {Caldas}}, \bibinfo {author} {\bibfnamefont {N.}~\bibnamefont {Marwan}},\ and\ \bibinfo {author} {\bibfnamefont {J.}~\bibnamefont {Kurths}},\ }\bibfield  {title} {\bibinfo {title} {Recurrence-based characterization of stickiness in {H}amiltonian systems},\ }in\ \href {https://doi.org/10.1007/978-3-031-91062-3_6} {\emph {\bibinfo {booktitle} {Recurrence Plots and Their Quantifications: Methodological Breakthroughs and Interdisciplinary Discoveries}}},\ \bibinfo {editor} {edited by\ \bibinfo {editor} {\bibfnamefont {Y.}~\bibnamefont {Hirata}}, \bibinfo {editor} {\bibfnamefont {M.}~\bibnamefont {Shiro}}, \bibinfo {editor} {\bibfnamefont {M.}~\bibnamefont {Fukino}}, \bibinfo {editor} {\bibfnamefont {C.~L.}\ \bibnamefont {Webber~Jr.}}, \bibinfo {editor} {\bibfnamefont {K.}~\bibnamefont {Aihara}},\ and\ \bibinfo {editor} {\bibfnamefont {N.}~\bibnamefont {Marwan}}}\ (\bibinfo  {publisher} {Springer Nature Switzerland},\ \bibinfo {year} {2025})\ pp.\ \bibinfo {pages} {95--108}\BibitemShut {NoStop}%
\bibitem [{\citenamefont {Gabrick}\ \emph {et~al.}(2023)\citenamefont {Gabrick}, \citenamefont {Sales}, \citenamefont {Sayari}, \citenamefont {Trobia}, \citenamefont {Lenzi}, \citenamefont {Borges}, \citenamefont {{Szezech Jr.}}, \citenamefont {Iarosz}, \citenamefont {Viana}, \citenamefont {Caldas},\ and\ \citenamefont {Batista}}]{Gabrick2023}%
  \BibitemOpen
  \bibfield  {author} {\bibinfo {author} {\bibfnamefont {E.~C.}\ \bibnamefont {Gabrick}}, \bibinfo {author} {\bibfnamefont {M.~R.}\ \bibnamefont {Sales}}, \bibinfo {author} {\bibfnamefont {E.}~\bibnamefont {Sayari}}, \bibinfo {author} {\bibfnamefont {J.}~\bibnamefont {Trobia}}, \bibinfo {author} {\bibfnamefont {E.~K.}\ \bibnamefont {Lenzi}}, \bibinfo {author} {\bibfnamefont {F.~S.}\ \bibnamefont {Borges}}, \bibinfo {author} {\bibfnamefont {J.~D.}\ \bibnamefont {{Szezech Jr.}}}, \bibinfo {author} {\bibfnamefont {K.~C.}\ \bibnamefont {Iarosz}}, \bibinfo {author} {\bibfnamefont {R.~L.}\ \bibnamefont {Viana}}, \bibinfo {author} {\bibfnamefont {I.~L.}\ \bibnamefont {Caldas}},\ and\ \bibinfo {author} {\bibfnamefont {A.~M.}\ \bibnamefont {Batista}},\ }\bibfield  {title} {\bibinfo {title} {Fractional dynamics and recurrence analysis in cancer model},\ }\href {https://doi.org/10.1007/s13538-023-01359-w} {\bibfield  {journal} {\bibinfo  {journal} {Brazilian Journal of Physics}\ }\textbf {\bibinfo {volume} {53}},\ \bibinfo {pages} {145} (\bibinfo {year} {2023})}\BibitemShut {NoStop}%
\bibitem [{\citenamefont {{Hénon}}\ and\ \citenamefont {{Heiles}}(1964)}]{henonheiles}%
  \BibitemOpen
  \bibfield  {author} {\bibinfo {author} {\bibfnamefont {M.}~\bibnamefont {{Hénon}}}\ and\ \bibinfo {author} {\bibfnamefont {C.}~\bibnamefont {{Heiles}}},\ }\bibfield  {title} {\bibinfo {title} {The applicability of the third integral of motion: Some numerical experiments},\ }\href {https://doi.org/10.1086/109234} {\bibfield  {journal} {\bibinfo  {journal} {Astronomical Journal}\ }\textbf {\bibinfo {volume} {69}},\ \bibinfo {pages} {73} (\bibinfo {year} {1964})}\BibitemShut {NoStop}%
\bibitem [{\citenamefont {Blesa}\ \emph {et~al.}(2012)\citenamefont {Blesa}, \citenamefont {Seoane}, \citenamefont {Barrio},\ and\ \citenamefont {Sanju\'{a}n}}]{Blesa2012}%
  \BibitemOpen
  \bibfield  {author} {\bibinfo {author} {\bibfnamefont {F.}~\bibnamefont {Blesa}}, \bibinfo {author} {\bibfnamefont {J.~M.}\ \bibnamefont {Seoane}}, \bibinfo {author} {\bibfnamefont {R.}~\bibnamefont {Barrio}},\ and\ \bibinfo {author} {\bibfnamefont {M.~A.~F.}\ \bibnamefont {Sanju\'{a}n}},\ }\bibfield  {title} {\bibinfo {title} {To escape or not to escape, that is the question — perturbing the {H}énon–{H}eiles {H}amiltonian},\ }\href {https://doi.org/10.1142/S0218127412300108} {\bibfield  {journal} {\bibinfo  {journal} {International Journal of Bifurcation and Chaos}\ }\textbf {\bibinfo {volume} {22}},\ \bibinfo {pages} {1230010} (\bibinfo {year} {2012})}\BibitemShut {NoStop}%
\bibitem [{\citenamefont {Zotos}(2017)}]{Zotos2017}%
  \BibitemOpen
  \bibfield  {author} {\bibinfo {author} {\bibfnamefont {E.~E.}\ \bibnamefont {Zotos}},\ }\bibfield  {title} {\bibinfo {title} {An overview of the escape dynamics in the {H}{\'e}non--{H}eiles {H}amiltonian system},\ }\href {https://doi.org/10.1007/s11012-017-0647-8} {\bibfield  {journal} {\bibinfo  {journal} {Meccanica}\ }\textbf {\bibinfo {volume} {52}},\ \bibinfo {pages} {2615} (\bibinfo {year} {2017})}\BibitemShut {NoStop}%
\bibitem [{\citenamefont {Vallejo}\ \emph {et~al.}(2025)\citenamefont {Vallejo}, \citenamefont {Nieto}, \citenamefont {Seoane},\ and\ \citenamefont {Sanju\'an}}]{Vallejo2025}%
  \BibitemOpen
  \bibfield  {author} {\bibinfo {author} {\bibfnamefont {J.~C.}\ \bibnamefont {Vallejo}}, \bibinfo {author} {\bibfnamefont {A.~R.}\ \bibnamefont {Nieto}}, \bibinfo {author} {\bibfnamefont {J.~M.}\ \bibnamefont {Seoane}},\ and\ \bibinfo {author} {\bibfnamefont {M.~A.~F.}\ \bibnamefont {Sanju\'an}},\ }\bibfield  {title} {\bibinfo {title} {Fast and slow escapes in forced chaotic scattering: The {N}ewtonian and the relativistic regimes},\ }\href {https://doi.org/10.1103/PhysRevE.111.024212} {\bibfield  {journal} {\bibinfo  {journal} {Phys. Rev. E}\ }\textbf {\bibinfo {volume} {111}},\ \bibinfo {pages} {024212} (\bibinfo {year} {2025})}\BibitemShut {NoStop}%
\bibitem [{\citenamefont {Yoshida}(1990)}]{Yoshida1990}%
  \BibitemOpen
  \bibfield  {author} {\bibinfo {author} {\bibfnamefont {H.}~\bibnamefont {Yoshida}},\ }\bibfield  {title} {\bibinfo {title} {Construction of higher order symplectic integrators},\ }\href {https://doi.org/https://doi.org/10.1016/0375-9601(90)90092-3} {\bibfield  {journal} {\bibinfo  {journal} {Physics Letters A}\ }\textbf {\bibinfo {volume} {150}},\ \bibinfo {pages} {262} (\bibinfo {year} {1990})}\BibitemShut {NoStop}%
\bibitem [{\citenamefont {Verlet}(1967)}]{verlet1967}%
  \BibitemOpen
  \bibfield  {author} {\bibinfo {author} {\bibfnamefont {L.}~\bibnamefont {Verlet}},\ }\bibfield  {title} {\bibinfo {title} {Computer "experiments" on classical fluids. i. thermodynamical properties of {L}ennard-{J}ones molecules},\ }\href {https://doi.org/10.1103/PhysRev.159.98} {\bibfield  {journal} {\bibinfo  {journal} {Phys. Rev.}\ }\textbf {\bibinfo {volume} {159}},\ \bibinfo {pages} {98} (\bibinfo {year} {1967})}\BibitemShut {NoStop}%
\bibitem [{\citenamefont {Zaslavsky}(2002)}]{ZASLAVSKY2002}%
  \BibitemOpen
  \bibfield  {author} {\bibinfo {author} {\bibfnamefont {G.~M.}\ \bibnamefont {Zaslavsky}},\ }\bibfield  {title} {\bibinfo {title} {Chaos, fractional kinetics, and anomalous transport},\ }\href {https://doi.org/https://doi.org/10.1016/S0370-1573(02)00331-9} {\bibfield  {journal} {\bibinfo  {journal} {Physics Reports}\ }\textbf {\bibinfo {volume} {371}},\ \bibinfo {pages} {461} (\bibinfo {year} {2002})}\BibitemShut {NoStop}%
\bibitem [{\citenamefont {Kraemer}\ \emph {et~al.}(2018)\citenamefont {Kraemer}, \citenamefont {Donner}, \citenamefont {Heitzig},\ and\ \citenamefont {Marwan}}]{Kraemer2018}%
  \BibitemOpen
  \bibfield  {author} {\bibinfo {author} {\bibfnamefont {K.~H.}\ \bibnamefont {Kraemer}}, \bibinfo {author} {\bibfnamefont {R.~V.}\ \bibnamefont {Donner}}, \bibinfo {author} {\bibfnamefont {J.}~\bibnamefont {Heitzig}},\ and\ \bibinfo {author} {\bibfnamefont {N.}~\bibnamefont {Marwan}},\ }\bibfield  {title} {\bibinfo {title} {{Recurrence threshold selection for obtaining robust recurrence characteristics in different embedding dimensions}},\ }\href {https://doi.org/10.1063/1.5024914} {\bibfield  {journal} {\bibinfo  {journal} {Chaos: An Interdisciplinary Journal of Nonlinear Science}\ }\textbf {\bibinfo {volume} {28}},\ \bibinfo {pages} {085720} (\bibinfo {year} {2018})}\BibitemShut {NoStop}%
\bibitem [{\citenamefont {Zbilut}\ \emph {et~al.}(2002)\citenamefont {Zbilut}, \citenamefont {Zaldivar-Comenges},\ and\ \citenamefont {Strozzi}}]{Zbilut2002}%
  \BibitemOpen
  \bibfield  {author} {\bibinfo {author} {\bibfnamefont {J.~P.}\ \bibnamefont {Zbilut}}, \bibinfo {author} {\bibfnamefont {J.-M.}\ \bibnamefont {Zaldivar-Comenges}},\ and\ \bibinfo {author} {\bibfnamefont {F.}~\bibnamefont {Strozzi}},\ }\bibfield  {title} {\bibinfo {title} {Recurrence quantification based {L}iapunov exponents for monitoring divergence in experimental data},\ }\href {https://doi.org/https://doi.org/10.1016/S0375-9601(02)00436-X} {\bibfield  {journal} {\bibinfo  {journal} {Physics Letters A}\ }\textbf {\bibinfo {volume} {297}},\ \bibinfo {pages} {173} (\bibinfo {year} {2002})}\BibitemShut {NoStop}%
\bibitem [{\citenamefont {Thiel}\ \emph {et~al.}(2002)\citenamefont {Thiel}, \citenamefont {{Carmen Romano}}, \citenamefont {Kurths}, \citenamefont {Meucci}, \citenamefont {Allaria},\ and\ \citenamefont {Arecchi}}]{Thiel2002}%
  \BibitemOpen
  \bibfield  {author} {\bibinfo {author} {\bibfnamefont {M.}~\bibnamefont {Thiel}}, \bibinfo {author} {\bibfnamefont {M.}~\bibnamefont {{Carmen Romano}}}, \bibinfo {author} {\bibfnamefont {J.}~\bibnamefont {Kurths}}, \bibinfo {author} {\bibfnamefont {R.}~\bibnamefont {Meucci}}, \bibinfo {author} {\bibfnamefont {E.}~\bibnamefont {Allaria}},\ and\ \bibinfo {author} {\bibfnamefont {F.~T.}\ \bibnamefont {Arecchi}},\ }\bibfield  {title} {\bibinfo {title} {Influence of observational noise on the recurrence quantification analysis},\ }\href {https://doi.org/https://doi.org/10.1016/S0167-2789(02)00586-9} {\bibfield  {journal} {\bibinfo  {journal} {Physica D: Nonlinear Phenomena}\ }\textbf {\bibinfo {volume} {171}},\ \bibinfo {pages} {138} (\bibinfo {year} {2002})}\BibitemShut {NoStop}%
\bibitem [{\citenamefont {Schinkel}\ \emph {et~al.}(2008)\citenamefont {Schinkel}, \citenamefont {Dimigen},\ and\ \citenamefont {Marwan}}]{Schinkel2008}%
  \BibitemOpen
  \bibfield  {author} {\bibinfo {author} {\bibfnamefont {S.}~\bibnamefont {Schinkel}}, \bibinfo {author} {\bibfnamefont {O.}~\bibnamefont {Dimigen}},\ and\ \bibinfo {author} {\bibfnamefont {N.}~\bibnamefont {Marwan}},\ }\bibfield  {title} {\bibinfo {title} {Selection of recurrence threshold for signal detection},\ }\href {https://doi.org/10.1140/epjst/e2008-00833-5} {\bibfield  {journal} {\bibinfo  {journal} {The European Physical Journal Special Topics}\ }\textbf {\bibinfo {volume} {164}},\ \bibinfo {pages} {45} (\bibinfo {year} {2008})}\BibitemShut {NoStop}%
\bibitem [{\citenamefont {Ngamga}\ \emph {et~al.}(2007)\citenamefont {Ngamga}, \citenamefont {Nandi}, \citenamefont {Ramaswamy}, \citenamefont {Romano}, \citenamefont {Thiel},\ and\ \citenamefont {Kurths}}]{Ngamga2007}%
  \BibitemOpen
  \bibfield  {author} {\bibinfo {author} {\bibfnamefont {E.~J.}\ \bibnamefont {Ngamga}}, \bibinfo {author} {\bibfnamefont {A.}~\bibnamefont {Nandi}}, \bibinfo {author} {\bibfnamefont {R.}~\bibnamefont {Ramaswamy}}, \bibinfo {author} {\bibfnamefont {M.~C.}\ \bibnamefont {Romano}}, \bibinfo {author} {\bibfnamefont {M.}~\bibnamefont {Thiel}},\ and\ \bibinfo {author} {\bibfnamefont {J.}~\bibnamefont {Kurths}},\ }\bibfield  {title} {\bibinfo {title} {Recurrence analysis of strange nonchaotic dynamics},\ }\href {https://doi.org/10.1103/PhysRevE.75.036222} {\bibfield  {journal} {\bibinfo  {journal} {Phys. Rev. E}\ }\textbf {\bibinfo {volume} {75}},\ \bibinfo {pages} {036222} (\bibinfo {year} {2007})}\BibitemShut {NoStop}%
\bibitem [{\citenamefont {Ngamga}\ \emph {et~al.}(2008)\citenamefont {Ngamga}, \citenamefont {Buscarino}, \citenamefont {Frasca}, \citenamefont {Fortuna}, \citenamefont {Prasad},\ and\ \citenamefont {Kurths}}]{Ngamga2008}%
  \BibitemOpen
  \bibfield  {author} {\bibinfo {author} {\bibfnamefont {E.~J.}\ \bibnamefont {Ngamga}}, \bibinfo {author} {\bibfnamefont {A.}~\bibnamefont {Buscarino}}, \bibinfo {author} {\bibfnamefont {M.}~\bibnamefont {Frasca}}, \bibinfo {author} {\bibfnamefont {L.}~\bibnamefont {Fortuna}}, \bibinfo {author} {\bibfnamefont {A.}~\bibnamefont {Prasad}},\ and\ \bibinfo {author} {\bibfnamefont {J.}~\bibnamefont {Kurths}},\ }\bibfield  {title} {\bibinfo {title} {{Recurrence analysis of strange nonchaotic dynamics in driven excitable systems}},\ }\href {https://doi.org/10.1063/1.2897312} {\bibfield  {journal} {\bibinfo  {journal} {Chaos: An Interdisciplinary Journal of Nonlinear Science}\ }\textbf {\bibinfo {volume} {18}},\ \bibinfo {pages} {013128} (\bibinfo {year} {2008})}\BibitemShut {NoStop}%
\bibitem [{\citenamefont {Marwan}(2008)}]{Marwan2008}%
  \BibitemOpen
  \bibfield  {author} {\bibinfo {author} {\bibfnamefont {N.}~\bibnamefont {Marwan}},\ }\bibfield  {title} {\bibinfo {title} {A historical review of recurrence plots},\ }\href {https://doi.org/10.1140/epjst/e2008-00829-1} {\bibfield  {journal} {\bibinfo  {journal} {The European Physical Journal Special Topics}\ }\textbf {\bibinfo {volume} {164}},\ \bibinfo {pages} {3} (\bibinfo {year} {2008})}\BibitemShut {NoStop}%
\bibitem [{\citenamefont {Marwan}(2025)}]{Marwan2025}%
  \BibitemOpen
  \bibfield  {author} {\bibinfo {author} {\bibfnamefont {N.}~\bibnamefont {Marwan}},\ }\bibfield  {title} {\bibinfo {title} {A bibliographic view on recurrence plots and recurrence quantification analyses},\ }in\ \href@noop {} {\emph {\bibinfo {booktitle} {Recurrence Plots and Their Quantifications: Methodological Breakthroughs and Interdisciplinary Discoveries}}},\ \bibinfo {editor} {edited by\ \bibinfo {editor} {\bibfnamefont {Y.}~\bibnamefont {Hirata}}, \bibinfo {editor} {\bibfnamefont {M.}~\bibnamefont {Shiro}}, \bibinfo {editor} {\bibfnamefont {M.}~\bibnamefont {Fukino}}, \bibinfo {editor} {\bibfnamefont {C.~L.}\ \bibnamefont {Webber~Jr.}}, \bibinfo {editor} {\bibfnamefont {K.}~\bibnamefont {Aihara}},\ and\ \bibinfo {editor} {\bibfnamefont {N.}~\bibnamefont {Marwan}}}\ (\bibinfo  {publisher} {Springer Nature Switzerland},\ \bibinfo {address} {Cham},\ \bibinfo {year} {2025})\ pp.\ \bibinfo {pages} {1--27}\BibitemShut {NoStop}%
\bibitem [{\citenamefont {Marwan}(2026)}]{Marwan2026}%
  \BibitemOpen
  \bibfield  {author} {\bibinfo {author} {\bibfnamefont {N.}~\bibnamefont {Marwan}},\ }\bibfield  {title} {\bibinfo {title} {Energy-efficient recurrence quantification analysis},\ }\bibfield  {journal} {\bibinfo  {journal} {The European Physical Journal Special Topics}\ }\href {https://doi.org/10.1140/epjs/s11734-025-02121-w} {10.1140/epjs/s11734-025-02121-w} (\bibinfo {year} {2026})\BibitemShut {NoStop}%
\bibitem [{\citenamefont {Ngamga}\ \emph {et~al.}(2012)\citenamefont {Ngamga}, \citenamefont {Senthilkumar}, \citenamefont {Prasad}, \citenamefont {Parmananda}, \citenamefont {Marwan},\ and\ \citenamefont {Kurths}}]{Ngamga2012}%
  \BibitemOpen
  \bibfield  {author} {\bibinfo {author} {\bibfnamefont {E.~J.}\ \bibnamefont {Ngamga}}, \bibinfo {author} {\bibfnamefont {D.~V.}\ \bibnamefont {Senthilkumar}}, \bibinfo {author} {\bibfnamefont {A.}~\bibnamefont {Prasad}}, \bibinfo {author} {\bibfnamefont {P.}~\bibnamefont {Parmananda}}, \bibinfo {author} {\bibfnamefont {N.}~\bibnamefont {Marwan}},\ and\ \bibinfo {author} {\bibfnamefont {J.}~\bibnamefont {Kurths}},\ }\bibfield  {title} {\bibinfo {title} {Distinguishing dynamics using recurrence-time statistics},\ }\href {https://doi.org/10.1103/PhysRevE.85.026217} {\bibfield  {journal} {\bibinfo  {journal} {Phys. Rev. E}\ }\textbf {\bibinfo {volume} {85}},\ \bibinfo {pages} {026217} (\bibinfo {year} {2012})}\BibitemShut {NoStop}%
\bibitem [{\citenamefont {Kraemer}\ and\ \citenamefont {Marwan}(2019)}]{Kraemer2019}%
  \BibitemOpen
  \bibfield  {author} {\bibinfo {author} {\bibfnamefont {K.~H.}\ \bibnamefont {Kraemer}}\ and\ \bibinfo {author} {\bibfnamefont {N.}~\bibnamefont {Marwan}},\ }\bibfield  {title} {\bibinfo {title} {Border effect corrections for diagonal line based recurrence quantification analysis measures},\ }\href {https://doi.org/https://doi.org/10.1016/j.physleta.2019.125977} {\bibfield  {journal} {\bibinfo  {journal} {Physics Letters A}\ }\textbf {\bibinfo {volume} {383}},\ \bibinfo {pages} {125977} (\bibinfo {year} {2019})}\BibitemShut {NoStop}%
\bibitem [{\citenamefont {Sales}(2023)}]{SalesThesis2023}%
  \BibitemOpen
  \bibfield  {author} {\bibinfo {author} {\bibfnamefont {M.~R.}\ \bibnamefont {Sales}},\ }\emph {\bibinfo {title} {Dynamical aspects of {H}amiltonian systems: chaos, stickiness, and recurrence plots}},\ \href {https://ri.uepg.br/handle/123456789/4183} {Ph.D. thesis},\ \bibinfo  {school} {Universidade Estadual de Ponta Grossa}, \bibinfo {address} {Ponta Grossa, PR, Brazil} (\bibinfo {year} {2023})\BibitemShut {NoStop}%
\bibitem [{\citenamefont {Altmann}\ \emph {et~al.}(2004)\citenamefont {Altmann}, \citenamefont {da~Silva},\ and\ \citenamefont {Caldas}}]{Altmann2004}%
  \BibitemOpen
  \bibfield  {author} {\bibinfo {author} {\bibfnamefont {E.~G.}\ \bibnamefont {Altmann}}, \bibinfo {author} {\bibfnamefont {E.~C.}\ \bibnamefont {da~Silva}},\ and\ \bibinfo {author} {\bibfnamefont {I.~L.}\ \bibnamefont {Caldas}},\ }\bibfield  {title} {\bibinfo {title} {Recurrence time statistics for finite size intervals},\ }\href {https://doi.org/10.1063/1.1795491} {\bibfield  {journal} {\bibinfo  {journal} {Chaos: An Interdisciplinary Journal of Nonlinear Science}\ }\textbf {\bibinfo {volume} {14}},\ \bibinfo {pages} {975} (\bibinfo {year} {2004})}\BibitemShut {NoStop}%
\bibitem [{\citenamefont {Altmann}\ and\ \citenamefont {Kantz}(2005)}]{Altmann2005b}%
  \BibitemOpen
  \bibfield  {author} {\bibinfo {author} {\bibfnamefont {E.~G.}\ \bibnamefont {Altmann}}\ and\ \bibinfo {author} {\bibfnamefont {H.}~\bibnamefont {Kantz}},\ }\bibfield  {title} {\bibinfo {title} {Recurrence time analysis, long-term correlations, and extreme events},\ }\href {https://doi.org/10.1103/PhysRevE.71.056106} {\bibfield  {journal} {\bibinfo  {journal} {Phys. Rev. E}\ }\textbf {\bibinfo {volume} {71}},\ \bibinfo {pages} {056106} (\bibinfo {year} {2005})}\BibitemShut {NoStop}%
\bibitem [{\citenamefont {Skokos}(2001)}]{Skokos2001}%
  \BibitemOpen
  \bibfield  {author} {\bibinfo {author} {\bibfnamefont {C.}~\bibnamefont {Skokos}},\ }\bibfield  {title} {\bibinfo {title} {Alignment indices: a new, simple method for determining the ordered or chaotic nature of orbits},\ }\href {https://doi.org/10.1088/0305-4470/34/47/309} {\bibfield  {journal} {\bibinfo  {journal} {Journal of Physics A: Mathematical and General}\ }\textbf {\bibinfo {volume} {34}},\ \bibinfo {pages} {10029} (\bibinfo {year} {2001})}\BibitemShut {NoStop}%
\bibitem [{\citenamefont {Skokos}\ \emph {et~al.}(2004)\citenamefont {Skokos}, \citenamefont {Antonopoulos}, \citenamefont {Bountis},\ and\ \citenamefont {Vrahatis}}]{Skokos2004}%
  \BibitemOpen
  \bibfield  {author} {\bibinfo {author} {\bibfnamefont {C.}~\bibnamefont {Skokos}}, \bibinfo {author} {\bibfnamefont {C.}~\bibnamefont {Antonopoulos}}, \bibinfo {author} {\bibfnamefont {T.~C.}\ \bibnamefont {Bountis}},\ and\ \bibinfo {author} {\bibfnamefont {M.~N.}\ \bibnamefont {Vrahatis}},\ }\bibfield  {title} {\bibinfo {title} {Detecting order and chaos in {H}amiltonian systems by the {SALI} method},\ }\href {https://doi.org/10.1088/0305-4470/37/24/006} {\bibfield  {journal} {\bibinfo  {journal} {Journal of Physics A: Mathematical and General}\ }\textbf {\bibinfo {volume} {37}},\ \bibinfo {pages} {6269} (\bibinfo {year} {2004})}\BibitemShut {NoStop}%
\bibitem [{\citenamefont {Skokos}\ \emph {et~al.}(2007)\citenamefont {Skokos}, \citenamefont {Bountis},\ and\ \citenamefont {Antonopoulos}}]{Skokos2007}%
  \BibitemOpen
  \bibfield  {author} {\bibinfo {author} {\bibfnamefont {C.}~\bibnamefont {Skokos}}, \bibinfo {author} {\bibfnamefont {T.~C.}\ \bibnamefont {Bountis}},\ and\ \bibinfo {author} {\bibfnamefont {C.}~\bibnamefont {Antonopoulos}},\ }\bibfield  {title} {\bibinfo {title} {Geometrical properties of local dynamics in {H}amiltonian systems: The {G}eneralized {A}lignment {I}ndex ({GALI}) method},\ }\href {https://doi.org/https://doi.org/10.1016/j.physd.2007.04.004} {\bibfield  {journal} {\bibinfo  {journal} {Physica D: Nonlinear Phenomena}\ }\textbf {\bibinfo {volume} {231}},\ \bibinfo {pages} {30} (\bibinfo {year} {2007})}\BibitemShut {NoStop}%
\bibitem [{\citenamefont {Skokos}\ \emph {et~al.}(2008)\citenamefont {Skokos}, \citenamefont {Bountis},\ and\ \citenamefont {Antonopoulos}}]{Skokos2008}%
  \BibitemOpen
  \bibfield  {author} {\bibinfo {author} {\bibfnamefont {C.}~\bibnamefont {Skokos}}, \bibinfo {author} {\bibfnamefont {T.}~\bibnamefont {Bountis}},\ and\ \bibinfo {author} {\bibfnamefont {C.}~\bibnamefont {Antonopoulos}},\ }\bibfield  {title} {\bibinfo {title} {Detecting chaos, determining the dimensions of tori and predicting slow diffusion in {F}ermi--{P}asta--{U}lam lattices by the {G}eneralized {A}lignment {I}ndex method},\ }\href {https://doi.org/10.1140/epjst/e2008-00844-2} {\bibfield  {journal} {\bibinfo  {journal} {The European Physical Journal Special Topics}\ }\textbf {\bibinfo {volume} {165}},\ \bibinfo {pages} {5} (\bibinfo {year} {2008})}\BibitemShut {NoStop}%
\bibitem [{\citenamefont {Manos}\ \emph {et~al.}(2012)\citenamefont {Manos}, \citenamefont {Skokos},\ and\ \citenamefont {Antonopoulos}}]{Manos2012}%
  \BibitemOpen
  \bibfield  {author} {\bibinfo {author} {\bibfnamefont {T.}~\bibnamefont {Manos}}, \bibinfo {author} {\bibfnamefont {C.}~\bibnamefont {Skokos}},\ and\ \bibinfo {author} {\bibfnamefont {C.}~\bibnamefont {Antonopoulos}},\ }\bibfield  {title} {\bibinfo {title} {Probing the local dynamics of periodic orbits by the {G}eneralized {A}lignment {I}ndex ({GALI}) method},\ }\href {https://doi.org/10.1142/S0218127412502185} {\bibfield  {journal} {\bibinfo  {journal} {International Journal of Bifurcation and Chaos}\ }\textbf {\bibinfo {volume} {22}},\ \bibinfo {pages} {1250218} (\bibinfo {year} {2012})}\BibitemShut {NoStop}%
\bibitem [{\citenamefont {Skokos}\ and\ \citenamefont {Manos}(2016)}]{Skokos2016}%
  \BibitemOpen
  \bibfield  {author} {\bibinfo {author} {\bibfnamefont {C.~H.}\ \bibnamefont {Skokos}}\ and\ \bibinfo {author} {\bibfnamefont {T.}~\bibnamefont {Manos}},\ }\bibinfo {title} {The smaller ({SALI}) and the generalized ({GALI}) alignment indices: Efficient methods of chaos detection},\ in\ \href {https://doi.org/10.1007/978-3-662-48410-4_5} {\emph {\bibinfo {booktitle} {Chaos Detection and Predictability}}}\ (\bibinfo  {publisher} {Springer Berlin Heidelberg},\ \bibinfo {address} {Berlin, Heidelberg},\ \bibinfo {year} {2016})\ pp.\ \bibinfo {pages} {129--181}\BibitemShut {NoStop}%
\bibitem [{\citenamefont {Sales}\ \emph {et~al.}(2026)\citenamefont {Sales}, \citenamefont {Leonel},\ and\ \citenamefont {Antonopoulos}}]{Sales2026}%
  \BibitemOpen
  \bibfield  {author} {\bibinfo {author} {\bibfnamefont {M.~R.}\ \bibnamefont {Sales}}, \bibinfo {author} {\bibfnamefont {E.~D.}\ \bibnamefont {Leonel}},\ and\ \bibinfo {author} {\bibfnamefont {C.~G.}\ \bibnamefont {Antonopoulos}},\ }\bibfield  {title} {\bibinfo {title} {On the behavior of {L}inear {D}ependence, {S}maller, and {G}eneralized {A}lignment {I}ndices in discrete and continuous chaotic systems},\ }\href {https://doi.org/https://doi.org/10.1016/j.chaos.2026.117884} {\bibfield  {journal} {\bibinfo  {journal} {Chaos, Solitons \& Fractals}\ }\textbf {\bibinfo {volume} {205}},\ \bibinfo {pages} {117884} (\bibinfo {year} {2026})}\BibitemShut {NoStop}%
\bibitem [{\citenamefont {Bazzani}\ \emph {et~al.}(2023)\citenamefont {Bazzani}, \citenamefont {Giovannozzi}, \citenamefont {Montanari},\ and\ \citenamefont {Turchetti}}]{Bazzani2023}%
  \BibitemOpen
  \bibfield  {author} {\bibinfo {author} {\bibfnamefont {A.}~\bibnamefont {Bazzani}}, \bibinfo {author} {\bibfnamefont {M.}~\bibnamefont {Giovannozzi}}, \bibinfo {author} {\bibfnamefont {C.~E.}\ \bibnamefont {Montanari}},\ and\ \bibinfo {author} {\bibfnamefont {G.}~\bibnamefont {Turchetti}},\ }\bibfield  {title} {\bibinfo {title} {Performance analysis of indicators of chaos for nonlinear dynamical systems},\ }\href {https://doi.org/10.1103/PhysRevE.107.064209} {\bibfield  {journal} {\bibinfo  {journal} {Phys. Rev. E}\ }\textbf {\bibinfo {volume} {107}},\ \bibinfo {pages} {064209} (\bibinfo {year} {2023})}\BibitemShut {NoStop}%
\bibitem [{\citenamefont {Escande}(1985)}]{Escande1985}%
  \BibitemOpen
  \bibfield  {author} {\bibinfo {author} {\bibfnamefont {D.~F.}\ \bibnamefont {Escande}},\ }\bibfield  {title} {\bibinfo {title} {Stochasticity in classical {H}amiltonian systems: Universal aspects},\ }\href {https://doi.org/https://doi.org/10.1016/0370-1573(85)90019-5} {\bibfield  {journal} {\bibinfo  {journal} {Physics Reports}\ }\textbf {\bibinfo {volume} {121}},\ \bibinfo {pages} {165} (\bibinfo {year} {1985})}\BibitemShut {NoStop}%
\bibitem [{\citenamefont {Contopoulos}\ and\ \citenamefont {Magnenat}(1985)}]{Contopoulos1985}%
  \BibitemOpen
  \bibfield  {author} {\bibinfo {author} {\bibfnamefont {G.}~\bibnamefont {Contopoulos}}\ and\ \bibinfo {author} {\bibfnamefont {P.}~\bibnamefont {Magnenat}},\ }\bibfield  {title} {\bibinfo {title} {Simple three-dimensional periodic orbits in a galactic-type potential},\ }\href {https://doi.org/10.1007/BF01261627} {\bibfield  {journal} {\bibinfo  {journal} {Celestial mechanics}\ }\textbf {\bibinfo {volume} {37}},\ \bibinfo {pages} {387} (\bibinfo {year} {1985})}\BibitemShut {NoStop}%
\bibitem [{\citenamefont {Grassberger}\ and\ \citenamefont {Procaccia}(1983)}]{Grassberger1983}%
  \BibitemOpen
  \bibfield  {author} {\bibinfo {author} {\bibfnamefont {P.}~\bibnamefont {Grassberger}}\ and\ \bibinfo {author} {\bibfnamefont {I.}~\bibnamefont {Procaccia}},\ }\bibfield  {title} {\bibinfo {title} {Measuring the strangeness of strange attractors},\ }\href {https://doi.org/10.1016/0167-2789(83)90298-1} {\bibfield  {journal} {\bibinfo  {journal} {Physica D: Nonlinear Phenomena}\ }\textbf {\bibinfo {volume} {9}},\ \bibinfo {pages} {189} (\bibinfo {year} {1983})}\BibitemShut {NoStop}%
\bibitem [{\citenamefont {Bandt}\ and\ \citenamefont {Pompe}(2002)}]{Bandt2002}%
  \BibitemOpen
  \bibfield  {author} {\bibinfo {author} {\bibfnamefont {C.}~\bibnamefont {Bandt}}\ and\ \bibinfo {author} {\bibfnamefont {B.}~\bibnamefont {Pompe}},\ }\bibfield  {title} {\bibinfo {title} {Permutation entropy: A natural complexity measure for time series},\ }\href {https://doi.org/10.1103/PhysRevLett.88.174102} {\bibfield  {journal} {\bibinfo  {journal} {Phys. Rev. Lett.}\ }\textbf {\bibinfo {volume} {88}},\ \bibinfo {pages} {174102} (\bibinfo {year} {2002})}\BibitemShut {NoStop}%
\bibitem [{\citenamefont {Sales}\ \emph {et~al.}(2025)\citenamefont {Sales}, \citenamefont {{de Souza}}, \citenamefont {Borin}, \citenamefont {Mugnaine}, \citenamefont {Szezech}, \citenamefont {Viana}, \citenamefont {Caldas}, \citenamefont {Leonel},\ and\ \citenamefont {Antonopoulos}}]{pynamicalsys}%
  \BibitemOpen
  \bibfield  {author} {\bibinfo {author} {\bibfnamefont {M.~R.}\ \bibnamefont {Sales}}, \bibinfo {author} {\bibfnamefont {L.~C.}\ \bibnamefont {{de Souza}}}, \bibinfo {author} {\bibfnamefont {D.}~\bibnamefont {Borin}}, \bibinfo {author} {\bibfnamefont {M.}~\bibnamefont {Mugnaine}}, \bibinfo {author} {\bibfnamefont {J.~D.}\ \bibnamefont {Szezech}}, \bibinfo {author} {\bibfnamefont {R.~L.}\ \bibnamefont {Viana}}, \bibinfo {author} {\bibfnamefont {I.~L.}\ \bibnamefont {Caldas}}, \bibinfo {author} {\bibfnamefont {E.~D.}\ \bibnamefont {Leonel}},\ and\ \bibinfo {author} {\bibfnamefont {C.~G.}\ \bibnamefont {Antonopoulos}},\ }\bibfield  {title} {\bibinfo {title} {pynamicalsys: A {P}ython toolkit for the analysis of dynamical systems},\ }\href {https://doi.org/https://doi.org/10.1016/j.chaos.2025.117269} {\bibfield  {journal} {\bibinfo  {journal} {Chaos, Solitons \& Fractals}\ }\textbf {\bibinfo {volume} {201}},\ \bibinfo {pages} {117269} (\bibinfo {year} {2025})}\BibitemShut {NoStop}%
\end{thebibliography}

%

\end{document}